\documentclass[sigconf]{acmart}
\AtBeginDocument{%
  }
\usepackage{subcaption}

\copyrightyear{2026}
\acmYear{2026}
\setcopyright{cc}
\setcctype{by}
\acmConference[CHI '26]{Proceedings of the 2026 CHI Conference on Human Factors in Computing Systems}{April 13--17, 2026}{Barcelona, Spain}
\acmBooktitle{Proceedings of the 2026 CHI Conference on Human Factors in Computing Systems (CHI '26), April 13--17, 2026, Barcelona, Spain}
\acmDOI{10.1145/3772318.3790358}
\acmISBN{979-8-4007-2278-3/2026/04}

\renewcommand\footnotetextcopyrightpermission[1]{}
\begin{document}

\title{From Overload to Convergence: Supporting Multi-Issue Human-AI Negotiation with Bayesian Visualization}


\author{Mehul Parmar}
\email{mehul.parmar@ait.asia}
\orcid{0000-0002-5311-6547}
\affiliation{%
  \institution{Asian Institute of Technology}
  \city{Bangkok}
  \country{Thailand}
}
\authornote{© Author 2026. This is the author's version of the work. It is posted here for your personal use. Not for redistribution. The definitive Version of Record was published in \textit{CHI '26}, \url{http://dx.doi.org/10.1145/3772318.3790358}.}

\author{Chaklam Silpasuwanchai}
\email{chaklam@ait.asia}
\orcid{0000-0002-8670-6893}
\affiliation{%
  \institution{Asian Institute of Technology}
  \city{Bangkok}
  \country{Thailand}
}

\renewcommand{\shortauthors}{Parmar et al.}

\begin{abstract}
As AI systems increasingly mediate negotiations, understanding how the number of negotiated issues impacts human performance is crucial for maintaining human agency.
We designed a human–AI negotiation case study in a realistic property rental scenario, varying the number of negotiated issues; empirical findings show that without support, performance stays stable up to three issues but declines as additional issues increase cognitive load.
To address this, we introduce a novel uncertainty-based visualization driven by Bayesian estimation of agreement probability. It shows how the space of mutually acceptable agreements narrows as negotiation progresses, helping users identify promising options.
In a within-subjects experiment (N=32), it improved human outcomes and efficiency, preserved human control, and avoided redistributing value.
Our findings surface practical limits on the complexity people can manage in human–AI negotiation, advance theory on human performance in complex negotiations, and offer validated design guidance for interactive systems.

\end{abstract}



\thanks{v2: Added Appendix \ref{sec:complete_prompts} (system prompts) and Appendix \ref{sec:experimental_payoff_matrices} (payoff matrices) in response to replication requests. Dataset independently available at https://doi.org/10.5281/zenodo.20545331}

\maketitle


\section{INTRODUCTION}
\label{sec:introduction}

    Imagine negotiating your next lease, flight upgrade, or supplier contract.  
    The party across the table is not a person—it is an AI system that can track multiple issues at once, process hundreds of counter-offers per second, and infer your preferences from your bidding patterns.
    While not yet ubiquitous, such systems are rapidly emerging—from Walmart’s Pactum AI saving millions in procurement \cite{van2022walmart} to Qatar Airways optimizing logistics in milliseconds \cite{qatar2024digital}, and legal-aid bots drafting settlements nightly \cite{westermann2023llmediator}. Yet, the missing piece is the \textit{human} in this loop: we lack systematic evidence on how humans fare when negotiating multiple issues against powerful automated agents that can evaluate countless trade-offs at once, potentially overwhelming our working memory and strategic reasoning.
    Consider a concrete example: negotiating a property rental where you must simultaneously balance monthly rent, security deposit, lease duration, and pet policy. You might accept higher rent for a pet-friendly longer lease. However, tracking these interdependent trade-offs and inferring the AI's preferences across countless combinations quickly exceeds human working memory. The AI can evaluate these trade-offs instantly, while you struggle to remember which combinations you've already explored.

    Prior negotiation research on the effects of \textbf{dimensionality} (the number of issues discussed simultaneously) in human–human contexts offers conflicting insights.
    One perspective posits that additional issues create opportunities for integrative trade‑offs, enhancing joint outcomes \cite{naquin2003agony,vanderschalk2010complex}. Another argues that increased dimensionality overwhelms human working memory, leading to suboptimal decisions \cite{arnold2021less,geiger2020more}. 
    Human-AI negotiation studies, however, have largely focused on single-issue or fixed multi-issue scenarios \cite{shen2024bargaining,jang2018more}, leaving the impact of varying dimensionality underexplored \cite{zhang2021negotiation,tsiakas2024unpacking}. 

    This gap is amplified by asymmetries: humans have limited cognitive capacity and difficulty modeling others' preferences, while AI can consider many possible trade-offs at once \cite{mell2020effects,arnold2021less}. Such disparities introduce interaction frictions that disadvantage humans and alter negotiation dynamics in ways unaddressed by prior work \cite{wang2024theory,geiger2020more,ward2024chatgpt}.
    To bridge these gaps, we investigate two research questions in a single-case study of property rental negotiations, where results generalize to other integrative domains but require validation.
    
    \textbf{RQ1:} In the absence of interface support, how does negotiating more issues at once affect human negotiators’ performance when interacting with AI negotiators?

    \textbf{RQ2:} Can an uncertainty-driven visualization tool reduce negotiation friction and improve human negotiators' performance across these levels of dimensionality?

    We developed a Decision Support tool that visualizes uncertainty about the opponent's acceptable zones. It consists of two widgets: a live heat-map of the agreement space and a progress panel showing convergence over time, that extend working memory. 
    Grounded in Bayesian inference (a way of updating beliefs as new offers appear), the tool models uncertainty about the AI's preferences, learning from each offer to guide users toward promising trade-offs without dictating their choices.
    We conducted a 2 (Interface: Baseline vs. Decision Support) $\times$ 4 (Dimensionality: 1, 3, 5, 7 issues) within-subjects experiment (N=32) in a realistic property rental scenario with integrative issues and asymmetric private payoffs.

    Key findings reveal a plateau-cliff effect in the baseline condition: human payoffs remain stable up to three issues but degrade sharply thereafter, confirming a bounded window within which humans can manage trade-offs effectively.
    The Decision Support tool mitigates this collapse, stabilizing payoffs and reducing cognitive load, particularly at higher dimensionality, while preserving neutral payoff distributions. 
    Subjective measures indicate reduced temporal demand and enhanced strategy satisfaction, despite longer interactions.

    This work makes three contributions to HCI and human-AI collaboration:
    \begin{itemize}
    \item \textbf{Empirical characterization of cognitive thresholds:} We delineate the plateau-cliff phenomenon in human-AI negotiation, extending theories of bounded rationality to multi-issue AI contexts and identifying dimensionality as a key source of friction.
    \item \textbf{Design of a Decision Support tool:} Our interface presents a reusable design pattern for cognitive offloading by visualizing uncertainty in two ways: a heat-map showing where to find agreement and a progress bar showing convergence over time.
    \item \textbf{Cognitive Harmony principle:} We contribute a design-oriented theoretical lens for augmentative interventions that reduce wasted effort without reallocating surplus, deriving principles for legitimate, agency-preserving AI augmentation.
    \end{itemize}
    These contributions inform the design of equitable human-AI systems, ensuring humans remain empowered partners in complex, AI-mediated decisions.

\section{RELATED WORK}
\label{sec:related_work}

    This section reviews prior work on cognitive limits and visualization based decision support in multi-issue human-AI negotiations, grounding our investigation of dimensionality thresholds (RQ1) and decision support design (RQ2).
    We define a negotiation \textbf{issue} as an agenda item requiring resolution to reach agreement \cite{bendahan2005negotiation}, excluding intangibles such as pride or saving face \cite{rubin1975social,geiger2020more}.
    Each issue comprises a set of \textbf{options}: the discrete choices available for that issue. Every \textit{option} has an associated \textbf{payoff}: the incentive a party receives for selecting that option.

    \subsection{Cognitive Limits and Dimensionality in Multi-Issue Negotiations}
    \label{sec:human_ai_negotiation}
        
        The landscape of human-AI negotiation has evolved from early rule-based systems \cite{filzmoser2010automated,demelo2011effect} to LLM-based agents capable of natural language dialogue and strategic reasoning \cite{lin2024decision, davidson2024evaluating, yang2020examining}. However, contemporary research reveals a persistent gap: while LLM agents demonstrate sophisticated capabilities \cite{davidson2024evaluating, shankar2024validates}, empirical studies typically examine single-issue contexts or fixed multi-issue scenarios without systematic manipulation of dimensionality \cite{zhang2021negotiation, jang2018more}.

        Scholars have called for systematic investigation of dimensionality's effects. Recent work in collaborative decision-making demonstrates that AI support must preserve human agency in high-stakes scenarios \cite{tsiakas2024unpacking, ward2024chatgpt}. 
        Previous attempts to quantify dimensionality's impact have yielded mixed results: some studies suggest that additional issues enhance integrative agreements and joint gains \cite{naquin2003agony, vanderschalk2010complex}, while others find added complexity overwhelms cognition, degrading outcomes \cite{arnold2021less, geiger2020more}.

        A recent meta-analysis of over 300 papers reports a sharp threshold at three issues in human-human bargaining, where additional issues degrade outcomes \cite{warsitzka2024expanding}; yet this does not address human-AI dynamics: where cognitive processing and strategic reasoning differ fundamentally \cite{wang2021mutual, mou2017media}; nor does it examine technological interventions. As AI agents proliferate in decision-making \cite{shen2024bargaining, ramachandran2022contract}, understanding dimensionality's interaction with human cognitive limits is essential.

        Evaluation of such systems requires attention to both objective and subjective outcomes. 
        Objective metrics include agreement rates, payoffs, total turns, and joint gains \cite{lai2007pareto, lai2009generic, sato2023preference}; while subjective metrics assess usability, cognitive load, confidence, and trust in AI partners \cite{amershi2019guidelines, zheng2022ux}.
        This dual framework reflects the HCI principle that effective human-AI systems must optimize both task performance and user experience \cite{yang2020examining, berthet_cognitive_2022}. Importantly, negotiation research distinguishes distributive contexts: zero-sum bargaining where one party's gain is another's loss \cite{lee2021comparing}, from integrative contexts, where multi-issue trade-offs allow for mutual gains \cite{embracing_complexity_psy_2023, warsitzka2024expanding}. While distributive tasks dominate early studies \cite{demelo2011effect}, integrative negotiations better capture real-world collaboration and are indispensable for examining dimensional complexity \cite{zhang2021negotiation, laubert2018disentangling}. Using integrative tasks ensures that performance differences reflect issue dimensionality rather than competitive dynamics, aligning with related experimental designs \cite{bosse2004experiments} and supporting our focus on bounded human-AI negotiation dynamics \cite{bazerman2000negotiation, robu2005modeling}.
        
    \subsection{Human-AI Negotiation Agents and Behavioral Asymmetries}

        Research on negotiation dimensionality highlights a core asymmetry: as the number of issues increases, human performance degrades due to cognitive overload \cite{arnold2021less}, while AI agents sustain consistent processing across higher-dimensional spaces \cite{mell2020effects}. This divergence reshapes negotiation dynamics. AI can systematically explore trade-off spaces to improve joint gains \cite{filzmoser2010automated}, yet humans face compounded burdens: managing complexity while modeling opaque AI reasoning \cite{gal2022multi}.
        Humans struggle with high-dimensional AI negotiations but can partially adapt over time \cite{mell2020effects}. 
        
        LLM-based agents add further variability: their behavior shifts under dimensional load \cite{davidson2024evaluating}, and advanced reasoning complicates anticipating moves. Humans rely on Theory of Mind to infer intentions \cite{wang2024theory}, yet human-to-human heuristics often fail with AI \cite{wang2021towards}. Without support, humans navigate unfamiliar AI strategies under heavy cognitive load, risking degraded payoffs and agency loss.

        LLMs like GPT-4 demonstrate strong negotiation competence \cite{davidson2024evaluating, jones2022capturing}, but require mitigation against biases: anchoring \cite{nguyen2024anchoring}, fairness-seeking \cite{davidson2024evaluating,amershi2019guidelines}, sycophancy \cite{ranaldi2023sycophantic}, and premature satisficing \cite{jones2022capturing}.
        Common safeguards include: (1) a utility-maximizing persona; (2) information asymmetry via exclusive AI access to its payoff matrix; (3) prompts avoiding middle-option defaults; and (4) iterative pilot refinement \cite{davidson2024evaluating}. 
        Generation parameters—temperature and response length—shape behavior, so fixing these across conditions isolates manipulated variables \cite{ward2024chatgpt, davidson2024evaluating}.

        Despite well-documented cognitive limits and AI asymmetries, no prior work empirically identifies dimensionality thresholds in human-AI negotiations or deploys adaptive tools that track and surface uncertainty about opponent preferences.
        
    \subsection{Visualization and Decision Support for Uncertainty in Complex Human-AI Interactions}
    \label{sec:support_tools_negotiation}
    
        Visualization and decision support mitigate performance degradation in complex negotiations. Graphical aids improve outcomes by helping users recognize beneficial exchanges \cite{gettinger2014far, kersten1999www}, while transparency and fairness shape user perceptions and outcomes \cite{holstein2022improving, cai2022human, wang2021designing}.

        However, existing interfaces face three critical challenges: (1) tracking evolving agreement spaces across issues, (2) synthesizing progress across dimensions, and (3) detecting inconsistent AI behavior \cite{zhan2024let}. These limitations degrade performance \cite{gero2020mental, wang2021towards}, reduce Theory of Mind accuracy \cite{westby2023collective}, and erode trust \cite{bansal2019beyond}. Current tools rely on descriptive methods—confidence bands or heuristic filtering—that visualize offers but lack frameworks for managing uncertainty about opponent preferences \cite{kersten2017heuristics, johnson2021comparing}.

        Consider a rental negotiation where a tenant proposes \$1,200 rent. The landlord must infer: does the tenant prioritize price, or are they flexible on cost if pets are allowed? This inference under uncertainty mirrors \textbf{Bayesian reasoning}: updating beliefs about another person's preferences in light of new information. Humans naturally engage in this process: after observing several offers emphasizing pet policies, the landlord revises their initial assumptions and infers the tenant values pet accommodation highly.
        However, when negotiations involve many issues simultaneously, tracking these evolving beliefs exceeds human cognitive capacity, especially against opaque AI agents \cite{westby2023collective,bansal2019beyond}.

        Bayesian methods formalize this intuition by integrating initial assumptions (priors) with observed evidence to produce updated beliefs. In contrast, \textbf{information entropy}: a measure of uncertainty or unpredictability in a distribution, quantifies remaining uncertainty. High entropy indicates many plausible preference structures remain; low entropy indicates preferences are nearly determined. Our work leverages them both: Bayesian updating maintains coherent beliefs across issues \cite{zeng1998bayesian}, while entropy-based visualizations direct attention to the most uncertain issues, helping users prioritize focus \cite{velloso2021probabilistic}.

        Despite Bayesian concepts appearing in HCI decision support \cite{westby2023collective,kim2019bayesian}, negotiation interfaces rarely leverage them, leaving a critical gap in managing opponent-model uncertainty under high dimensionality. Our visualization applies entropy-weighted cues to surface contentious issues, aiming to reduce cognitive friction and sustain human agency in human-AI negotiations.

\section{METHODOLOGY}
\label{sec:methodology}

    To address our two research questions—(RQ1) the impact of negotiation dimensionality on human-AI outcomes, and (RQ2) the effect of an entropy-driven Bayesian visualization tool—we conducted a 2×4 within-subjects experiment varying issue count (1, 3, 5, 7) and tool availability.

    \subsection{Design of Negotiation Tasks}
    \label{sec:design_negotiation_tasks}
        We designed a property rental negotiation scenario comprising 16 integrative issues, each with 7 distinct options, to systematically manipulate task dimensionality while maintaining ecological validity through participants' familiarity with rental contexts. 
        Below we discuss the key design tradeoffs and validation strategies employed to ensure experimental rigor.

        \subsubsection{Scenario Selection and Ecological Validity} We used a property rental negotiation (tenant vs. landlord) to keep the task realistic and leverage participants' domain familiarity. 
        Unlike abstract paradigms that can reduce ecological validity \cite{gero2020mental,lee2019virtual}, this setting preserves real-world relevance while maintaining experimental control. 
        Participants always played the tenant and the AI the landlord, exploiting the natural role asymmetry (tenants seek favorable terms; landlords maximize value) to mirror utility-driven objectives. 
        Keeping roles constant across dimensionality conditions isolated issue-count effects while preserving engagement and realistic decision-making \cite{kuang2024enhancing,lim2024exploring}.

        \begin{figure}[h]
            \centering
            \includegraphics[width=0.80\linewidth]{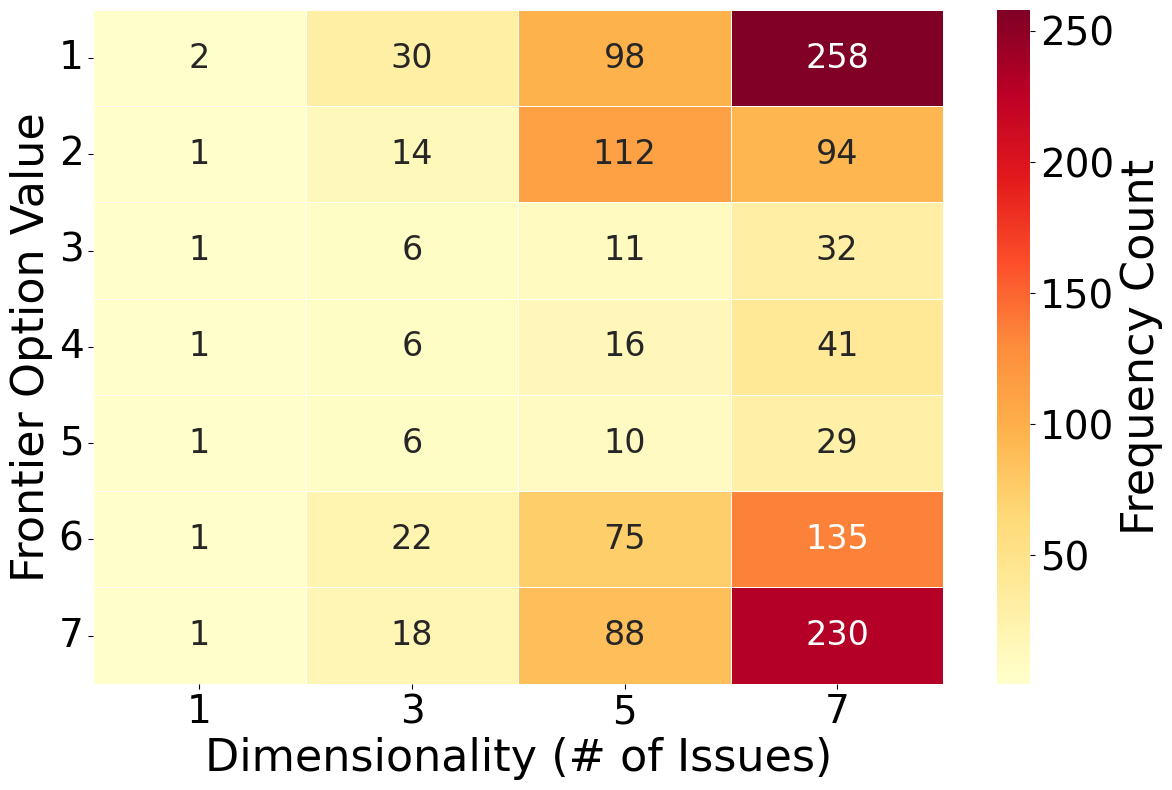}
            \caption{Heatmap of Pareto‑optimal placements: 2‑D distribution of joint‑optimum points across issues, showing dispersed optima that encourage exploratory search rather than midpoint convergence.}
            \label{fig:ch3-issue-option-design}
            \Description{A 2‑D heatmap (counts / density) plotting the locations of Pareto‑optimal (joint best‑outcome) option combinations derived from the task payoff matrices across the four issue conditions. Color intensity encodes the frequency of Pareto‑optimal points at each option index; axes show option index (horizontal) and issue or aggregated bin (vertical). The distribution demonstrates that Pareto optima are spread across the option space rather than concentrated at a single or central index, a deliberate design choice to promote exploratory behavior and avoid trivial midpoint strategies.}
        \end{figure}        

        \begin{table*}[h]
            \centering
            \caption{Payoff distribution for one issue used in our study (``Utilities Included''). The design places the joint optimum at a non-obvious position, distinct from either party's maximum or the central option, promoting exploration.}
            \label{tab:utilities-payoff}
            \begin{tabular}{l l c c c l}
                \toprule
                Sr. No. & Option & Human Payoff & AI Payoff & Joint Payoff & Remark \\
                \midrule
                1 & None & 5 & 95 & 100 & max AI (landlord) payoff \\
                2 & Internet only & 15 & 85 & 100 &  \\
                3 & Internet + Water & 30 & 80 & 110 &  \\
                4 & All except electric & 55 & 65 & 120 & middle option \\
                5 & All utilities & 80 & 50 & 130 &  \\
                6 & All utilities + cleaning & 100 & 35 & 135 & max joint utility \\
                7 & All inclusive & 110 & 20 & 130 & max human (tenant) payoff \\
                \bottomrule
            \end{tabular}
        \end{table*}      

        \subsubsection{Issue Design and Structural Choices} Within this property rental negotiation framework, we made three key structural decisions to ensure experimental control while maintaining realism. 
        First, we used only \textit{integrative issues} (issues where parties can make trade-offs across options to create mutual value, rather than purely distributive zero-sum conflicts), as this is essential for multi-issue negotiations to achieve optimal outcomes.
        Second, all issues were \textit{equally weighted}. Introducing weights raises cognitive load and lets the AI exploit complex trade-offs, which can mask dimensionality effects; equal weighting therefore isolates issue count as the independent variable and ensures fair, comparable conditions.
        Third, payoff structures were intentionally \textit{asymmetric} to encourage exploration and realistic trade-offs. 
        For each issue we varied which options maximize each party's utility, often placing optima at opposite ends of the option spectrum (Figure~\ref{fig:ch3-issue-option-design}) to avoid premature midpoint convergence and to surface integrative opportunities \cite{hancock2008i, jaidka2024takes}.

        Optima here are \textbf{Pareto‑optimal} options: outcomes where neither tenant nor landlord can improve their payoff without making the other worse off. These represent the theoretically ideal agreements that maximize total value creation.
        These options collectively define the \textit{Pareto frontier}, the full set of efficient trade-offs between competing objectives.
        The distance of joint outcomes from the Pareto frontier reveals how much value participants leave on the table, answering RQ1 on efficiency and RQ2 on tool benefit.      
        
        \subsubsection{Anti-triviality and hidden-information safeguards} 
        \label{subsubsec:anti_triviality_safeguards}
            To prevent trivial heuristics, payoff progressions (e.g., monotonic, plateaued, inverted) were de-correlated across the 16 issues and perturbed slightly so joint optima could not be inferred from a single private matrix. 
            For example, consider a negotiation over the ``Utilities Included'' issue (see Table~\ref{tab:utilities-payoff}). 
            The human's best option (option 7) yields 110 points but only 20 for the AI. Conversely, the AI's best option (option 1) yields 95 points but only 5 for the human. Yet the joint optimum (option 6) yields a combined 135 points. 
            Since neither party can infer this from their private payoffs alone, the design encourages exploration and integrative reasoning. Settling on the intuitive midpoint (option 4) yields only 120 points, leaving 15 points of value uncaptured. This design forces iterative preference discovery rather than enabling analytic shortcuts, ensuring that gains from the Decision Support condition (RQ2) reflect genuine cognitive leverage rather than task simplicity, while isolating dimensionality effects for RQ1.

            In sum, the task balances experimental control (integrative issues, equal weights, asymmetric private payoffs) with ecological validity. 
            Participants negotiated 1, 3, 5, or 7 issues sampled from a 16-issue, 7-option set; private payoff matrices preserve hidden information so we can isolate dimensionality effects and evaluate the Bayesian visualization (RQ2).

    \subsection{Dependent Variables}
    \label{sec:dependent_variables}
        Dependent variables were grouped as proxies for performance and for friction; some measures served both purposes.
        \subsubsection{Proxies for Performance}
        \label{sec:proxies_performance}
            These measures assessed negotiation outcomes and efficiency, drawing from prior work on multi-issue negotiation and choice overload in decision-making contexts \cite{hale2022preference}.
            \begin{itemize}
            \item Total Human Payoff: The total payoff for each participant, at the end of each negotiation session, calculated as a percentage of maximum possible payoff for that session. It indicates individual negotiation success.
            \item Joint payoff: Sum of human and AI payoffs at the final agreement. It indicates overall negotiation efficiency and value creation.
            \item Pareto proximity: Distance of the joint payoff to the Pareto frontier, measured as the average Euclidean distance across issues (measured in number of payoff units). It denotes how much value was left on the table after negotiation.
            \item Total turns to agreement: Number of turns taken to reach a final agreement (or timeout). It reflects negotiation efficiency and process dynamics.
            \item Concession Behavior:
            A concession occurs when, for a given issue, a player offers an option that yields them a lower payoff than their previous offer. We measured:
                \begin{itemize}
                \item Per-turn concession magnitude: The amount a player's payoff decreases when they make a concession on a turn.
                \item Number of concessions: Count of turns where a player made a positive concession.
                \item Average concession magnitude: Mean size of a player's concessions, computed only over turns with concessions.
                \end{itemize}
            \item Confidence: Self-reported confidence in negotiation outcome (7-point Likert scale). The question asked was: ``How confident are you that you achieved the best possible outcome in this negotiation?''
            \item Strategy Satisfaction: Satisfaction with negotiation strategy employed (7-point Likert scale). The question asked was: ``How satisfied are you with the negotiation strategy you used in this session?''
            \end{itemize}
        \subsubsection{Proxies for Friction}
        \label{sec:proxies_friction}
            These measures captured cognitive and interactional challenges, such as overload from high dimensionality.
            \begin{itemize}
            \item Total chat duration: Total elapsed time from the first message to the final submitted agreement (or session timeout), in seconds. It reflects overall negotiation effort and pacing.
            \item Backtracking Frequency: Count of turns where the participant returns to an earlier proposal, indicating iterative re-consideration.
            \item Average First Keystroke Time: Elapsed time from receipt of the partner's message to the participant's first keystroke, averaged across all negotiation turns and measured in seconds. This metric indexes response‑initiation (deliberation) time; we exclude the first turn's latency because it is confounded with task reading time.
            \item Sequence entropy (a measure of uncertainty in choice patterns): Operationalized as Shannon entropy, it measures the variability of a participant's proposals. High entropy reflects wide exploration, while low entropy indicates a focused strategy (e.g., making small, repeated concessions on one issue).

            For $m$ issues we compute per-issue entropy $H_k$ and report the average:
            \begin{equation}
                	\text{Sequence Entropy} = \frac{1}{m} \sum_{k=1}^{m} H_k
            \end{equation}                
            \item Cognitive Load: NASA-TLX scale \cite{kosch2023survey}. Assessed mental effort required for each negotiation.
            \item User Experience: System Usability Scale (SUS) for visual tool conditions \cite{lewis2018system}.
            \end{itemize}
            At each negotiation \textbf{turn} (defined as a human proposal followed by an AI counter-proposal) we recorded turn-level interaction logs: the option(s) proposed by each party, the timestamp of the participant's first keystroke after receiving the partner's message, and the timestamp of the submitted message. 
            These low-level logs were combined with the payoff matrices to derive the objective dependent variables described above.

    \subsection{AI Agent Configuration}
    \label{sec:ai_agent_configuration}
        We employed OpenAI's GPT-4 (API access, mid-2025) as our AI negotiation agent. To mitigate known LLM biases, we implemented several safeguards informed by prior work, including a utility-maximizing persona, information asymmetry, prompts to prevent anchoring, and pilot testing to curb cooperative drift. 
        The human participant always initiated the negotiation. The LLM operated with a constant temperature of 0.2 and a 128-token limit to ensure consistent behavior, in line with similar prior work \cite{davidson2024evaluating}.

        \subsubsection{Example Prompts}
        \label{sec:example_prompts}
            Representative prompt excerpts (supporting the unified safeguard layer) are provided for replication transparency:

            \begin{itemize}
            \item System prompt: ``You are an AI negotiator in a property rental scenario. 
            Your primary objective is to maximize your own utility score based on the provided payoff matrix. 
            Avoid defaulting to compromise or fairness-based solutions unless they demonstrably increase your score. 
            Do not anchor on middle options without strategic justification. 
            Evaluate each proposal based solely on its impact on your utility and respond with clear strategic reasoning.''

            \item User prompt: ``Based on your payoff matrix for the current negotiation issues, propose options that maximize your total utility score and provide explicit justification for your choices. 
            Consider potential trade-offs across all issues when making your proposal.''
            \end{itemize}

            \noindent These prompts are inspired by recent work on AI negotiations \cite{davidson2024evaluating}, and incorporate specific bias-mitigation techniques identified in contemporary LLM negotiation research \cite{jones2022capturing,amershi2019guidelines}, including explicit utility-maximization directives, anti-anchoring instructions, and requirements for strategic justification.

    \subsection{Decision Support Tool}
    \label{sec:decision_support_tool}     
        
        As reviewed in Section~\ref{sec:support_tools_negotiation}, existing interfaces face three persistent challenges in high-dimensional human-AI negotiation: (1) tracking evolving agreement spaces, (2) synthesizing progress across many dimensions, and (3) detecting when AI behavior is inconsistent or irrational. 
        These gaps can lead to suboptimal agreements, missed trade-offs, and reduced user trust in AI partners.

        To counter these, we adopt a Bayesian formulation because: (a) it provides principled uncertainty quantification, allowing users to distinguish stronger from weaker inferred AI preferences; (b) it incrementally integrates prior belief with new conversational evidence, preserving early strategic signals; (c) it supports adaptive re-weighting when behavior becomes erratic without discarding accumulated structure; and (d) it enforces mathematical coherence across issues rather than maintaining parallel heuristic rules.
        We note that some benefits of our visualization could be approximated by alternative approaches (e.g., frequentist confidence intervals or heuristic filtering); our selection of a Bayesian formulation is pragmatic—chosen for unified updating and explicit uncertainty representation—rather than a claim of methodological exclusivity. The parameters for this model were set based on pilot data and established design heuristics; we do not claim optimality.

        Our Bayesian dashboard pairs two widgets, each closing a distinct negotiation gap; the next sections detail their design, mathematical foundations, and user-facing benefits.

        \begin{figure*}[h]
            \centering
            \includegraphics[width=0.95\linewidth]{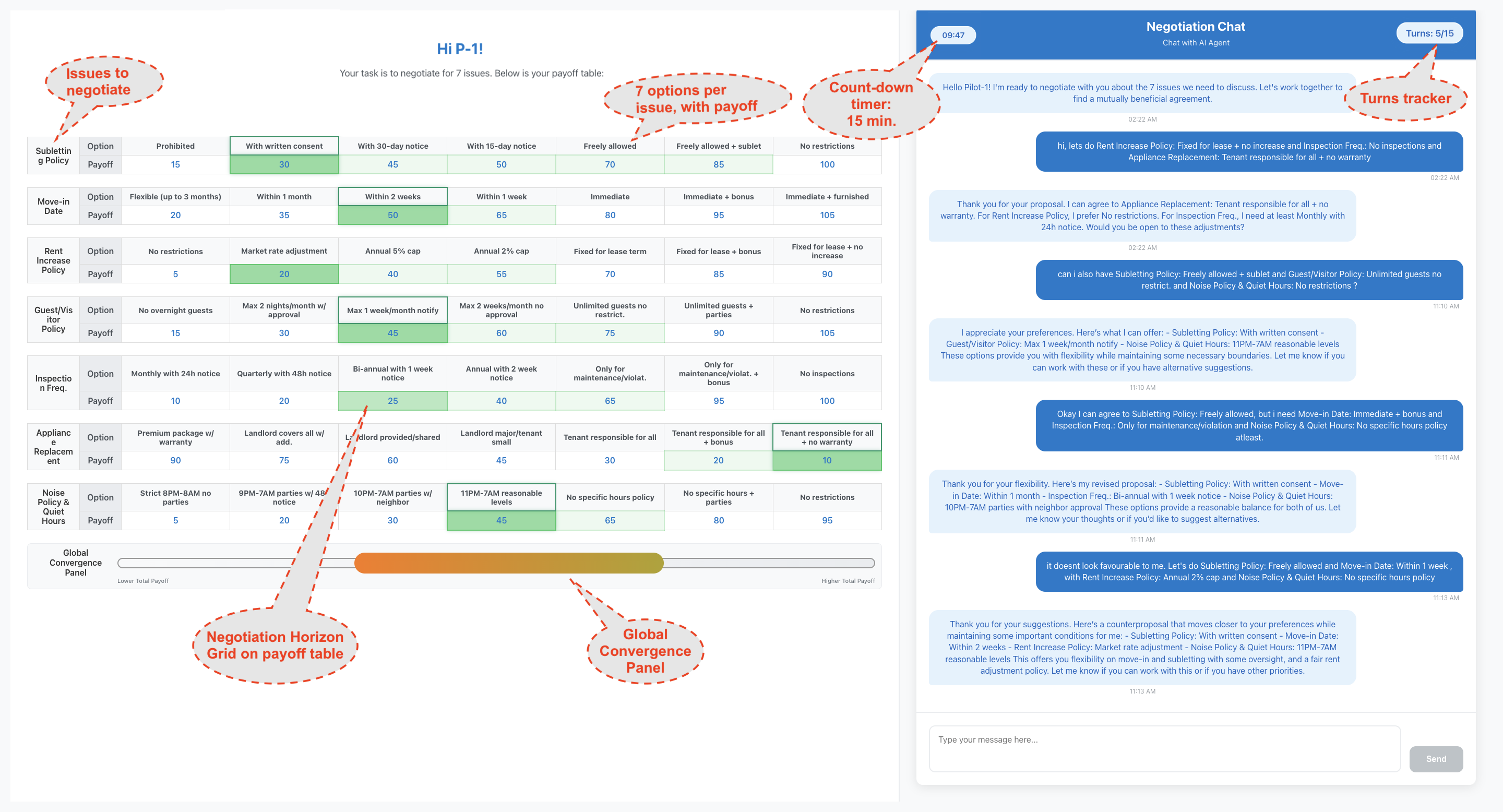}
            \caption{Sample conversation between a human participant and the AI agent, in the Decision Support condition, showing the visualization interface in action. Green bands of the Negotiation Horizon Grid indicate ZOPA, whereas the Global Convergence Panel shows overall convergence and favorability.}
            \label{fig:sample-conversation}
            \Description{Screenshot showing the negotiation chat alongside the Decision Support widgets: the Negotiation Horizon Grid (a multi-row 7-column grid with green ZOPA bands highlighting promising option ranges for each issue) and the Global Convergence Panel (a horizontal bar indicating overall convergence and favorability). The chat pane shows a recent human proposal and the AI counter-proposal; visual highlights indicate which options the system currently rates as most promising.}
        \end{figure*}    

        \subsubsection{Negotiation Horizon Grid}
        \label{sec:negotiation_horizon_grid}

            \paragraph{Problem Statement:} In high-dimensional human-AI negotiation, human negotiators experience severe information overload. They struggle to track the evolving set of mutually acceptable options, known as the Zone of Possible Agreement (\textbf{ZOPA}): the range of options both parties find acceptable. AI, in contrast, evaluates more trade-offs than humans can track. This makes it hard to detect inconsistent or irrational behavior.           
            
            \paragraph{Design Explanation:} The Negotiation Horizon Grid presents a $7 \times n$ grid layout. The grid is 7 columns wide (one for each option per issue) and $n$ rows deep, with $n$ corresponding to the number of issues in the current negotiation condition. Each row represents a negotiation issue. Each column corresponds to a discrete option value. The grid is overlaid with dynamic green ``Negotiation Horizon'' bands that visually represent the evolving ZOPA for each issue (Figure~\ref{fig:sample-conversation}).

            \paragraph{Mathematical Framework:} The Horizon Grid employs a dynamic Bayesian inference system. It continuously updates probability distributions over negotiation options based on agent behavior. The complete update equation integrates four core components:

            \begin{equation}
            \label{eq:bayesian-update}
            \begin{split}
                P(\text{option}_j | \text{evidence}_t) 
                &= \eta \times \mathcal{L}_{\text{base}}(\text{evidence}_t | \text{option}_j) \\ 
                &\quad \times B_{\text{ZOPA}}(j) \times P(\text{option}_j | \text{evidence}_{t-1}) \\
                &\quad \times W_{\text{consistency}}(\text{agent}) 
            \end{split}
            \end{equation}

            where $\eta$ is the normalization constant ensuring that the probabilities sum to 1, thus forming a proper probability mass function (PMF) at each step.
            
            \textit{In plain language, this equation combines four pieces of information to guess the AI's preferences: what the AI just proposed (Likelihood), what it has rejected in the past (ZOPA), our previous guess (Prior), and how consistently the AI is behaving (Weights). The result is then normalized to ensure all probabilities add up to 100\%.}
            
            Priors are initialized as uniform distributions over options at $t=0$, assuming no initial knowledge of AI preferences. 
            We discuss each of the four components of this Bayesian update equation below.

            \paragraph{Component 1 - Likelihood Function ($\mathcal{L}_{\text{base}}$):} 
            This component assigns higher probabilities to AI-proposed options and adjacent alternatives. It focuses user attention on promising negotiation areas.

            \begin{multline}
            \label{eq:likelihood-function}
            \mathcal{L}_{\text{base}}(\text{evidence}_t | \text{option}_j) = \\
            \begin{cases}
            0.8 \times C, & \text{if } j = j^* \text{ (direct proposal)} \\
            0.4, & \text{if } |j - j^*| \leq 1 \text{ (adjacent options)} \\
            0.1, & \text{otherwise (distant options)}
            \end{cases}
            \end{multline}

            where $j^*$ is the proposed option index and $C$ is the consistency score.
            
            \textit{This component makes the system pay more attention to options the AI recently proposed or those nearby. For example, if the AI proposes option 4 on an issue, the likelihood assigns 0.8$C$ to option 4, 0.4 to options 3 and 5, and 0.1 to all others.}

            \paragraph{Component 2 - ZOPA Boundary Constraints ($B_{\text{ZOPA}}$):} 
            This component filters out consistently rejected options. It focuses effort on realistic negotiation ranges. Boundary detection modulates likelihood based on AI proposal patterns:
            \begin{multline}
            \label{eq:zopa-boundary}
            B_{\text{ZOPA}}(j) = \\
            \begin{cases}
            1.0, & \text{if } j \in [\text{lowerLimit}, \text{upperLimit}] \\
            (1 - \text{boundaryConfidence}), & \text{if } j \notin [\text{lowerLimit}, \\
            & \quad \text{upperLimit}]
            \end{cases}
            \end{multline}
            Boundaries are computed from AI proposal history. Confidence is derived as:
            \begin{equation}
            \label{eq:boundary-confidence}
            \mathrm{boundaryConfidence} = 1 - \mathrm{normalized\_proposal\_variance}
            \end{equation}
            High proposal variance yields low confidence; tight clustering yields high confidence, down-weighting historically infeasible regions while preserving uncertainty.
            
            \textit{This component helps the system learn which options are ``off the table'' based on the AI's past offers.}

            \paragraph{Component 3 - Adaptive Weights ($W_{\text{consistency}}$):} 
            This component modulates trust in AI signals based on behavioral consistency. It applies agent-specific scaling factors that adjust probability weights:
            
            \begin{multline}
            \label{eq:adaptive-weights}
            W_{\text{consistency}}(\text{agent}) = \\
            \begin{cases}
            \min(1.0, 0.7 \times (1 + s_{\text{consistency}})), \hspace{1em} \text{ if agent = AI} \\
            \min(1.0, 0.3 \times (1 + |r_{\text{concession}}|)), \hspace{1em} \text{if agent = Human}
            \end{cases}
            \end{multline}

            where $s_{\text{consistency}} = 0.6 \times C_{\text{proposal}} + 0.4 \times C_{\text{temporal}}$. 
            
            This combines proposal variance and temporal slope analysis to detect erratic AI behavior. The component ensures the negotiation strategy prioritizes reliable, consistent agents. It down-weights erratic AI behavior, improving overall decision stability.
            
            \textit{Intuitively, this acts as a ``trust meter'': consistent AI behavior increases its influence on the visualization, whereas noisy behavior reduces it.}

            \paragraph{Component 4 - Visual Mapping ($P(\text{option}_j | \text{evidence}_{t-1})$):} 
            The final step translates the calculated probabilities into visual intensities (color saturation) on the grid. 
            
            The goal is to draw the user's attention to strategically valuable options without overwhelming them. High-intensity green highlights options that are both likely to be accepted by the AI and valuable to the user. Lower-intensity green indicates acceptable but less optimal alternatives.

            \begin{equation}
            \label{eq:visual-mapping}
            \begin{split}
            I_{i,j} = \begin{cases}
            \min\Big(0.6, P_{i,j} \times 2 \times \sqrt{\text{boundaryConfidence}_i} \\
            \qquad \times (1 + s_{\text{consistency}}) \times \xi_i\Big), \\
            \qquad \text{if } j \in \text{ZOPA}_i \text{ and } u_{h,i,j} \geq \tau_{\text{min},i} \\[1ex]
            \min(0.25, P_{i,j} \times 0.4), \\
            \qquad \text{if } \tau_{\text{min},i} \leq u_{h,i,j} \leq \tau_{\text{max},i} \\[1ex]
            0, \qquad \text{otherwise}
            \end{cases}
            \end{split}
            \end{equation}

            where $\text{ZOPA}_i =$
            \begin{equation}
            [\max(0, \lfloor\text{lowerLimit}_i\rfloor), \min(6, \lceil\text{upperLimit}_i\rceil)]
            \end{equation}
            
            The constants were derived from pilot data and design heuristics. The cap of \textbf{0.6} for high-intensity green prevents the grid from becoming oversaturated. This maintains visual clarity even with 7 issues. The multiplier of \textbf{2} amplifies the signal for promising options. The cap of \textbf{0.25} for low-intensity green ensures secondary options remain visually distinct. The multiplier of \textbf{0.4} for this tier preserves contrast. 
            
            The issue-specific scaling factor, $\xi_i$, allows for future adaptation based on issue importance. It was held at 1.0 in this study. Utility thresholds, $\tau_{\text{min},i}$ and $\tau_{\text{max},i}$, are derived from the user's own payoff table. They filter out options unacceptable to the user, ensuring the visualization only highlights mutually beneficial possibilities.
            
            \textit{In short, this equation paints the grid: options that are good for the user and likely to be accepted by the AI glow brighter green, while acceptable-but-worse options are a lighter green.}

            Overall, the Negotiation Horizon Grid transforms complex negotiation dynamics into intuitive visual feedback, helping users identify promising options and track agreement progress to manage cognitive load.
            A detailed worked example demonstrating the complete calculation pipeline is provided in Appendix~\ref{sec:worked-example-bayesian} for reproducibility.
            
        \subsubsection{Global Convergence Panel}
        \label{sec:global_convergence_panel}
            \paragraph{Problem Statement:} While tracking individual issues is challenging, an equally significant problem in multi-issue negotiation is perceiving how adjustments in one issue affect the overall negotiation outcome. Users often become fixated on optimizing single issues without understanding how these choices impact the global agreement potential.
            This myopic focus can lead to impasses in otherwise solvable negotiations. 
            Existing interfaces provide little support for understanding the holistic negotiation state, creating what negotiation theorists call ``the dimensionality trap'' \cite{gettinger2014far,dedreu2006motivated,zhang2021negotiation}.

            \begin{figure}[h]
                \centering
                \includegraphics[width=1\linewidth]{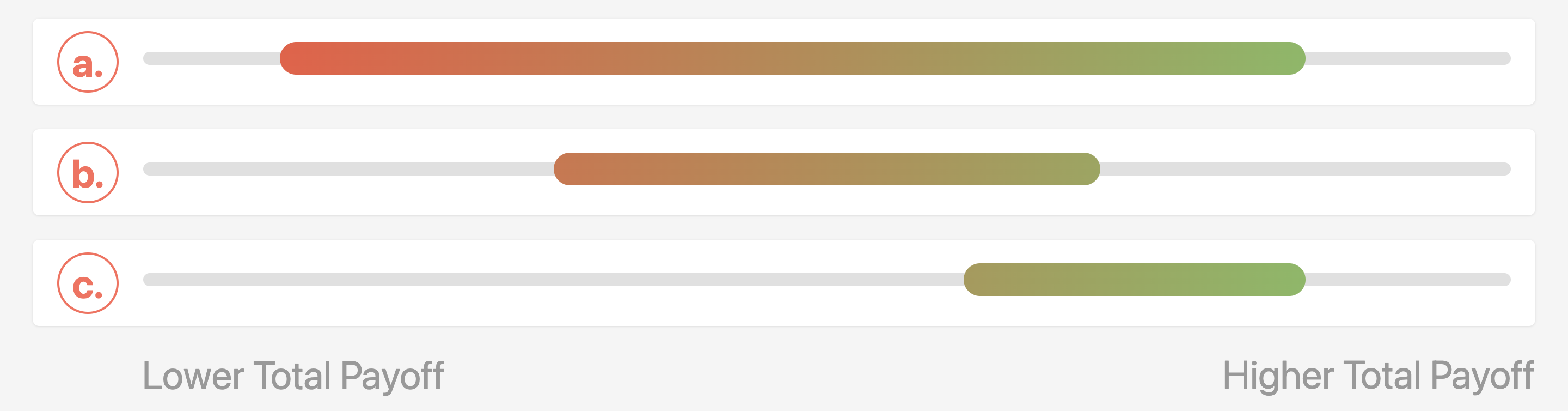}
                \caption{Panel (a) shows an early-stage negotiation where the wide colored zone indicates high uncertainty about likely agreement ranges; panel (b) shows a mid-stage negotiation with reduced uncertainty; panel (c) shows a potential late-stage convergence where the colored zone is much narrower (low uncertainty) and the ZOPA is increasingly favorable to the human participant.}
                \label{fig:convergence-panel}
                \Description{The Convergence Panel visualizes global agreement progress across three stages: (a) early-stage — a wide colored zone denotes high uncertainty about likely agreement ranges; (b) mid-stage — a reduced zone denotes declining uncertainty; (c) late-stage — a narrow zone shifted toward green denotes low uncertainty with the ZOPA favoring the human. The horizontal gradient bar therefore encodes both residual uncertainty (zone width) and strategic favorability (position on the red–amber–green spectrum).}
            \end{figure}   

        \begin{figure*}[t]
            \centering
            \includegraphics[width=0.8\linewidth]{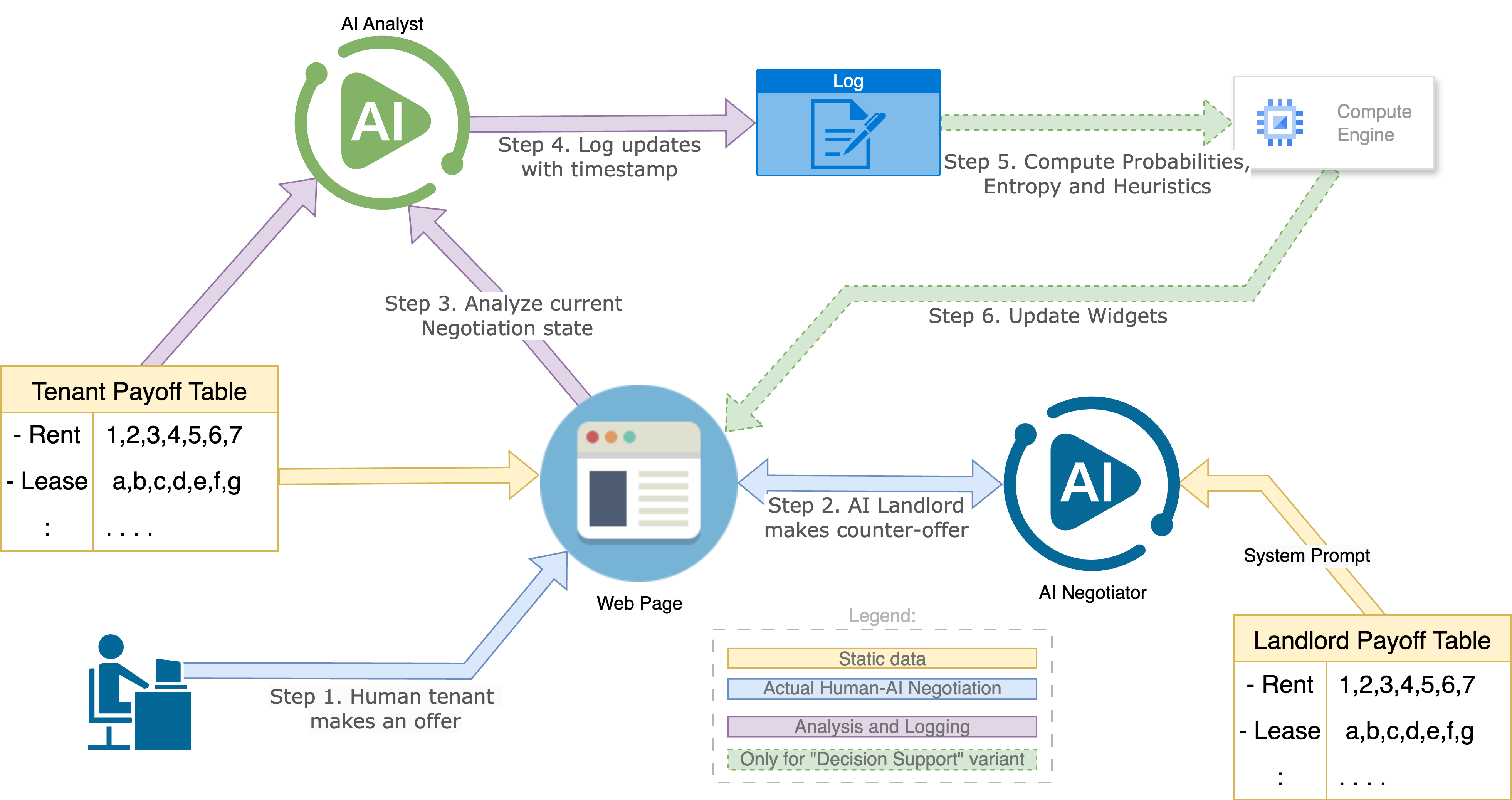}
            \caption{Overview of the negotiation chatbot system architecture.}
            \label{fig:system_design}
            \Description{System architecture diagram showing modules and data flow: user interface (human offers), AI Negotiator (generates counter-offers), AI Analyst (parses dialogue for metric computation), centralized logger, and the visualization widgets updated by the metric computation service. Arrows indicate real-time flow between components while labels denote information separation (private payoff matrices) and where Decision Support computations occur.}
        \end{figure*}  
        
            \paragraph{Design Explanation:} The Global Convergence Panel provides a holistic summary of the negotiation state across all issues. It features a horizontal bar with a gradient from red (high entropy, AI-favorable) through yellow (balanced) to green (low entropy, human-favorable).
            The width of the colored zone dynamically narrows as the negotiation progresses and uncertainty is reduced. Its position along the bar indicates whether the current mutually acceptable options (ZOPA) are more favorable to the human or less (see Figure~\ref{fig:convergence-panel}).

            \paragraph{Mathematical Framework:} The panel's visualization is driven by two core calculations. These determine its width and position, which are rendered in a synchronized manner.

            First, the bar's \textit{width} represents the scope of the negotiation space. It is calculated as the inverse of the convergence ratio:
            \begin{multline}
            \mathrm{width\_percentage} = \\
            \max((100 / 7), (1 - \mathrm{convergence\_ratio}) \times 100)
            \end{multline}
            where $\mathrm{convergence\_ratio}$ =
            \begin{equation}
            1 - (\mathrm{effective\_green\_cells} / \mathrm{total\_visible\_cells})
            \end{equation}
            A narrower bar signifies a smaller, more defined agreement space.
            
            Second, the bar's \textit{position} indicates which party the current agreement trajectory favors. It is determined by the weighted average payoff within the ZOPA:
            \begin{equation}
            \mathrm{weighted\_position} = \frac{\sum_{i=0}^{6} \mathrm{issue\_ZOPA\_average}_i \times (1/7)}{\sum_{i=0}^{6} \mathrm{issue\_max\_payoff}_i \times (1/7)} \times 100
            \end{equation}
            
            where $\mathrm{issue\_ZOPA\_average}_i$ is the mean payoff within the ZOPA range. The position is then mapped to a red-amber-green color gradient. This provides an at-a-glance summary of strategic favorability.
        
        Together, the Negotiation Horizon Grid and the Global Convergence Panel translate Bayesian inferences into actionable, cognitively efficient visual cues. These tools enable users to quickly identify promising options and monitor overall negotiation progress, supporting effective decision-making in complex, multi-issue human - AI interactions. We next describe the system architecture that implements these computations.

    \subsection{System Design}
    \label{sec:system_design}  
        
        The Negotiation Chatbot system is designed to facilitate and analyze human-AI negotiation interactions through an intuitive user interface.
        The underlying architecture (Figure~\ref{fig:system_design}) enforces strict information separation while enabling robust, real-time analysis of negotiation dynamics. The interaction proceeds as follows:

        \begin{enumerate}
        \item \textbf{User Offer:} The human participant, acting as the tenant, initiates each turn by submitting an offer via the web interface. The user's actions are informed only by their private payoff table, which is always visible.

        \item \textbf{AI Counter-Offer:} The AI Negotiator, acting as the landlord, receives the user's offer and generates a counter-offer. The AI's decisions are based solely on its private landlord payoff table, enforcing information asymmetry that mirrors real-world negotiations.

        \item \textbf{State Analysis:} Concurrently, an AI Analyst, which is an independent instance of GPT-4, parses the conversational exchange. This component has access only to the user's (tenant's) payoff table and structures the dialogue for downstream analysis and feedback, a technique known as LLM-Evaluator \cite{shankar2024validates}.

        \item \textbf{Interaction Logging:} All conversational turns, including offers, responses, and timestamps, are logged for real-time updates and post-hoc analysis.

        \item \textbf{Metric Computation (Decision Support Variant):} In the Decision Support condition, the system computes Bayesian probabilities, entropy, and other heuristics based on the logged data after each turn. These computations only use information available to the human participant.

        \item \textbf{Widget Update (Decision Support Variant):} The computed metrics are used to update the visualization widgets (Negotiation Horizon Grid and Global Convergence Panel) in near real-time ($\sim$1.4s latency), providing the user with Decision Support.
        \end{enumerate}

        This design ensures that every negotiation turn is analyzable, supporting rigorous experimental control for studying human-AI interaction in high-dimensional negotiation tasks.

    \subsection{Pilot Study}
    \label{sec:pilot_study}

        \begin{figure*}[h]
            \centering
            \includegraphics[width=0.90\linewidth]{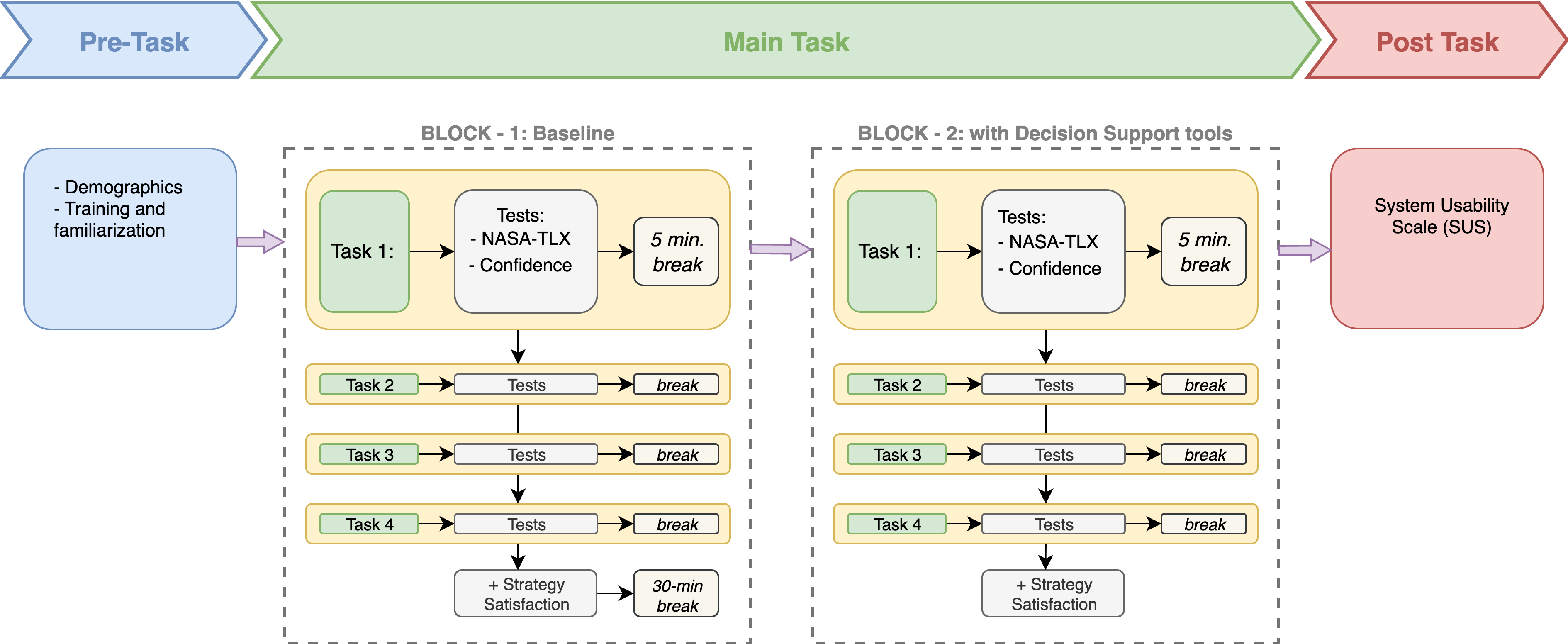}
            \caption{An overview of the Experimental Procedure.}
            \label{fig:methodology-procedure}
            \Description{Flowchart of the experiment timeline: informed consent and pre-task questionnaires, training on the baseline interface, Block 1 (four Baseline negotiation tasks with post-task NASA-TLX and brief ratings), a short break, Block 2 (four Decision Support tasks with identical post-task measures), and final SUS and debrief. The diagram emphasizes the fixed Interface order (Baseline then Decision Support) and per-block timing of approximately 30–45 minutes.}
        \end{figure*} 

        A rigorous 6-participant pilot study was conducted to refine the experimental protocol, validate instrumentation, and ensure task feasibility. The following key adjustments were made based on pilot data:
        \begin{itemize}
            \item Dimensionality Refinement: Pilot data revealed significant cognitive fatigue when dimensionality exceeded five issues. Consequently, we adopted a stepped set of 1, 3, 5, and 7 issues to ensure task tractability while capturing a meaningful range of complexity.
            \item Time and Round Caps: Analysis of pilot negotiations showed that agreements were typically reached within 13 rounds. To balance task completion with participant endurance, we instituted a cap of 15 rounds or 15 minutes, with mandatory 3-minute breaks between tasks to mitigate fatigue.
            \item Interface changes: (1) a single-screen layout (no scrolling) for up to seven issues to reduce visual scanning overhead; (2) an instruction for participants to briefly reflect before beginning to type so first-keystroke timing could be captured consistently; (3) short (0.5 s) update animations for band changes to make adjustments easier to follow; and (4) a copy-paste feature to transfer issue and option descriptions from the payoff table to the chatbox, minimizing typing effort.
            \item Parameter Sensitivity Checks: We tested sensitivity of key visualization parameters (Equation~\ref{eq:visual-mapping}). For instance, we varied the high-intensity saturation cap (from 0.5 to 0.7, final: 0.6) and the intensity multiplier (from 1.5 to 2.5, final: 2) to confirm that the visual distinction between promising and acceptable options remained clear and interpretable. This confirmed the chosen values were robust without implying a full optimization sweep.
            \item AI Analyst and Behavior Validation: Qualitative analysis of negotiation transcripts confirmed that the AI Negotiator's behavior remained utility-focused, with no systematic drift toward unprompted compromise, thus validating our bias mitigation safeguards. All AI Analyst inferences were manually verified by two independent coders and found correct.
            \item Instrumentation Integrity: The data logging pipeline was verified to capture complete, timestamped, turn-level interaction data without loss.
            \item Participant Exclusion: To prevent learning effects, all pilot participants were excluded from the main study.
        \end{itemize}
        These data-driven refinements finalized the experimental procedure and UI design prior to main data collection.
        
    \subsection{Participants}
    \label{sec:participants}

        We recruited 32 participants (15 male, 16 female, 1 non-binary; age 22–45, M=25.4, SD=6.2) from local universities and institutions, ensuring balanced gender representation \cite{geiger2020more}. 
        All were fluent English speakers with prior rental negotiation experience, basic computer literacy, and familiarity with chat interfaces and AI assistants. Educational backgrounds ranged from Bachelor's (n=23) to Master's (n=9), and negotiation experience varied from minimal to extensive. Participants received a base compensation equivalent to USD 10.

        \begin{table*}[h!]
          \centering
          \caption{Holm-adjusted post hoc pairwise comparisons of total human payoff across dimensionality levels in Baseline and Decision Support conditions. The table displays selected, non-exhaustive comparisons relevant to the main findings.}
          \label{tab:posthoc_baseline}
          \begin{tabular}{l ccc c ccc}
          \toprule
          & \multicolumn{3}{c}{\textbf{Baseline}} & & \multicolumn{3}{c}{\textbf{Decision Support}} \\
          \cmidrule{2-4} \cmidrule{6-8}
          \textbf{Comparison} & \textit{t}-value & \(p_{holm}\) & Cohen's \textit{d} & & \textit{t}-value & \(p_{holm}\) & Cohen's \textit{d} \\
          \midrule
          1 vs. 3 Issues & -2.32 & .054 & -0.668 & & -0.758 & 1.00 (n.s.) &-0.194 \\
          3 vs. 5 Issues & 7.54 & \textbf{$<$.001} & 1.12 & & 0.899 & 1.00 (n.s.) & 0.179 \\
          5 vs. 7 Issues & 6.82 & \textbf{$<$.001} & 0.967 & & 1.470 & 1.00 (n.s.) & 0.233 \\
          \bottomrule
          \end{tabular}%
        \end{table*}

    \subsection{Experimental Design Overview}
    \label{sec:experimental_design}

        We employed a 2 (Interface: Baseline, Decision Support) × 4 (Dimensionality: 1, 3, 5, 7) within-subjects design to examine the effects of negotiation dimensionality and visualization availability on human-AI negotiation performance.

        To isolate the impact of our visual tool, we adopted a fixed order for the \textit{Interface condition}: all participants completed the Baseline tasks before the Decision Support tasks. While counterbalancing is a standard method for mitigating order effects, we argue that for this specific research question, a fixed order provides a more rigorous design. 
        Exposing participants to the Decision Support tool first would likely contaminate their behavior in the subsequent Baseline condition; once participants have been exposed to the strategic insights provided by the visualizations, it is impossible for them to revert to a truly ``baseline'' state of mind \cite{cockburn2017effects, lazar2017research}. 
        This fixed progression also mirrors the natural adoption curve of new technologies, where users first engage with a basic system before adopting advanced features, thus enhancing ecological validity. 
        For the \textit{Dimensionality} factor, we used a Latin square design to counterbalance the order of the 1, 3, 5, and 7-issue conditions across participants, minimizing learning or fatigue effects related to task complexity.

        This within-subjects design maximizes statistical power and enables direct comparison of the Decision Support interface against Baseline across dimensionality levels while controlling for participant specific negotiation skill and strategy preferences. 
        All participants completed negotiations across eight crossed conditions: four Dimensionality levels × two Interface levels. 
        With N = 32, our design had sufficient sensitivity to detect medium to large effect sizes ($f \geq 0.25$) with 80\% power at $\alpha = 0.05$.

    \subsection{Procedure}
    \label{sec:procedure}
           
        \subsubsection{Session Overview}
        \label{sec:session_overview}
            Each participant attended a single 1-hour session in a controlled laboratory environment (Figure ~\ref{fig:methodology-procedure}). To ensure consistent conditions, sessions were individual. A standardized protocol, including instructions and timing, was used for all participants to ensure rigor.

        \subsubsection{Pre-Experiment Phase}
        \label{sec:pre_experiment_phase}
            After providing informed consent, participants completed questionnaires on demographics, prior negotiation experience, and technology familiarity. Data was logged with anonymized IDs. 
            Participants were then trained on the baseline interface with a mock scenario before being introduced to the main rental negotiation task, their role (tenant), and the AI's role (landlord). 
            Instructions emphasized utility maximization and the payoff structure. To incentivize strategic behavior, a performance-based bonus of up to \$30 equivalent was offered.

        \subsubsection{Main Experimental Sessions}
        \label{sec:main_experimental_sessions}
            The main experiment consisted of two blocks, each with four negotiation tasks of varying dimensionality (1, 3, 5, 7 issues):
            \textit{Block 1: Baseline}
            Participants first completed four negotiation tasks without the visual support tool. After each task, participants completed the NASA-TLX (cognitive load), and single-item Likert ratings for confidence. 
            At the end of the block, they rated their strategy satisfaction. This block lasted approximately 30-45 minutes.
            \textit{Block 2: Decision Support}
            After a short break, participants completed four additional negotiation tasks with the visual support tool enabled. The same post-task measures were collected after each task, and strategy satisfaction was rated at the end of the block. 
            This block also lasted approximately 30-45 minutes.

        \subsubsection{Post-Session Measures}
        \label{sec:post_session_measures}
            After both blocks, participants completed the SUS questionnaire. System logs were collected for objective performance analysis. 
            Statistical analysis was then used to compare outcomes across conditions.

\section{RESULTS}
\label{sec:results}

  \subsection{Overview}
  \label{sec:results_overview}
    This section reports results from our 2 × 4 within-subjects experiment. We first present manipulation and sanity checks, then report primary objective outcomes (payoffs, efficiency), followed by process metrics, subjective measures, and a summary of findings. For all repeated-measures ANOVAs, we checked the assumption of sphericity using Mauchly's test; where this assumption was violated, a Greenhouse–Geisser correction was applied to the degrees of freedom. 
    All AI model and visualization parameters (Eqs.~\ref{eq:bayesian-update} and \ref{eq:visual-mapping}) were fixed prior to data collection; no post-hoc optimization or parameter tuning was performed based on study outcomes.

  \subsection{Manipulation \& Sanity Checks}
  \label{sec:manipulation_checks}
    To ensure internal validity, we verified that counterbalancing dimensionality produced no systematic differences across order positions. A repeated-measures ANOVA with SequenceGroup (A–D) as a between-subjects factor showed no main effect of sequence on total payoff (\textit{F}(3,28) = 0.53, \textit{p} = .667), and no interactions with the main experimental factors (all \textit{p}'s $>$ .13). This confirms that learning or fatigue effects were not confounded with our manipulations. AI consistency checks indicated no invalid proposals across all sessions, and pilot validation confirmed consistent agent behavior. We manually verified the consistency of the AI Analyst by randomly checking conversation logs with the AI Analyst's interpretation, and found that the AI Analyst correctly extracted the proposals and counter-proposals from the ongoing conversation in real time.
    
  \subsection{Primary Objective Outcomes}
  \label{sec:primary_outcomes}

      \begin{figure*}[h]
        \centering
        \begin{subfigure}[b]{0.32\linewidth}
          \centering
          \includegraphics[width=\linewidth]{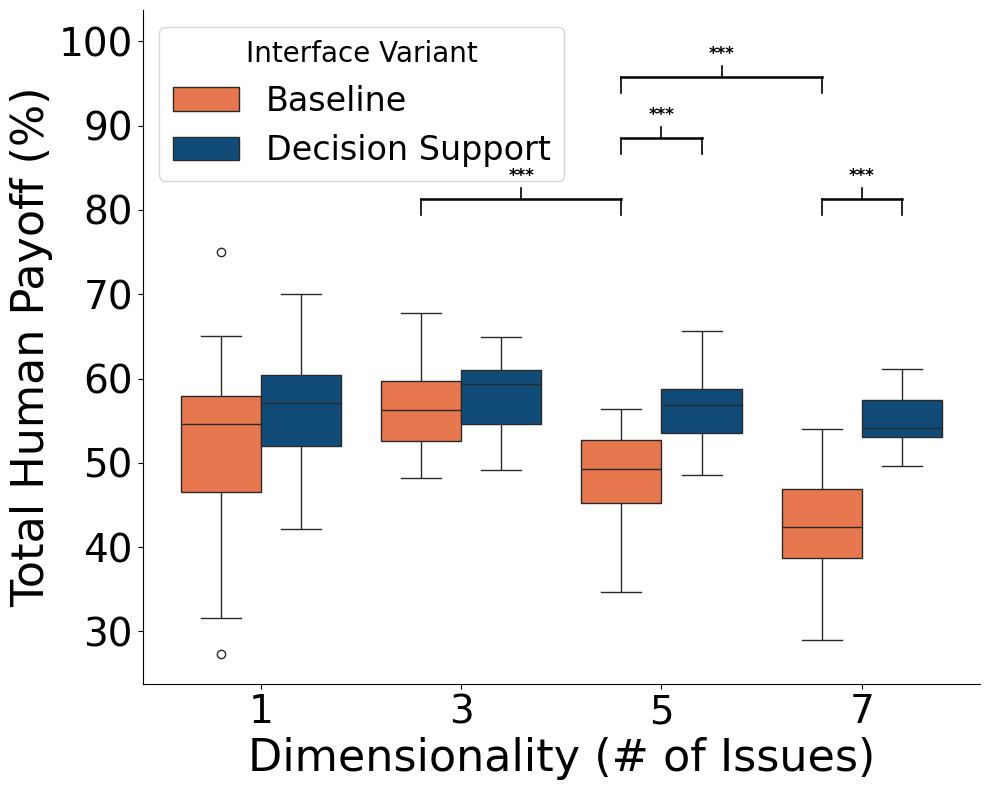}
          \caption{Total human payoff.}
          \label{fig:total-human-payoff}
        \end{subfigure}\hfill
        \begin{subfigure}[b]{0.32\linewidth}
          \centering
          \includegraphics[width=\linewidth]{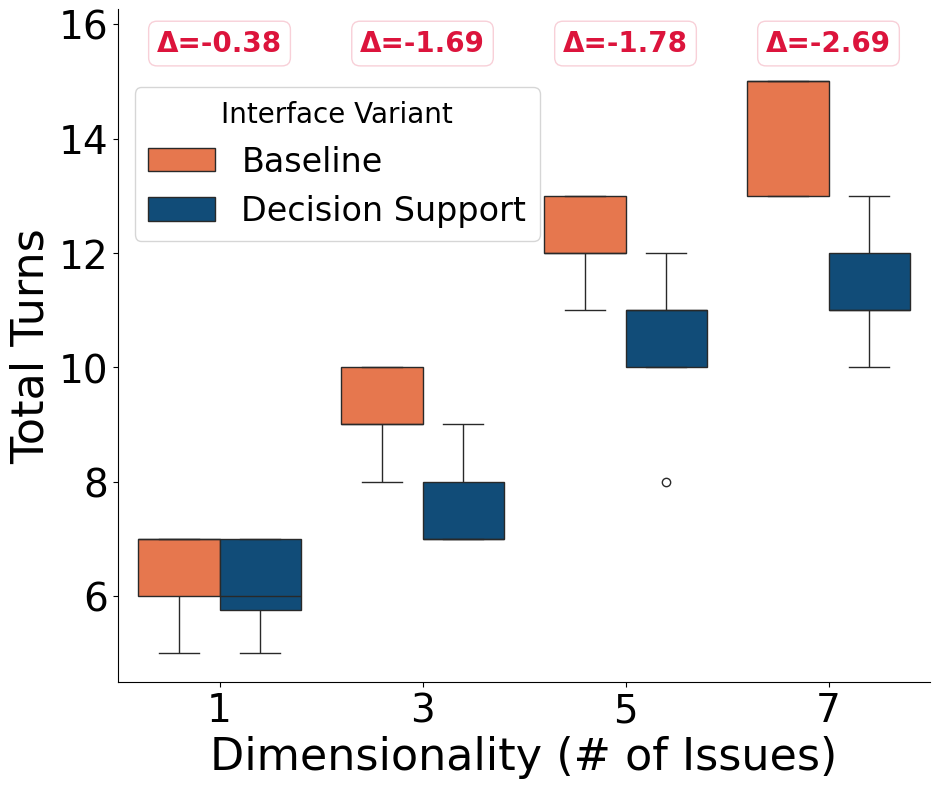}
          \caption{Total turns to agreement.}
          \label{fig:total-turns}
        \end{subfigure}\hfill
        \begin{subfigure}[b]{0.32\linewidth}
          \centering
          \includegraphics[width=\linewidth]{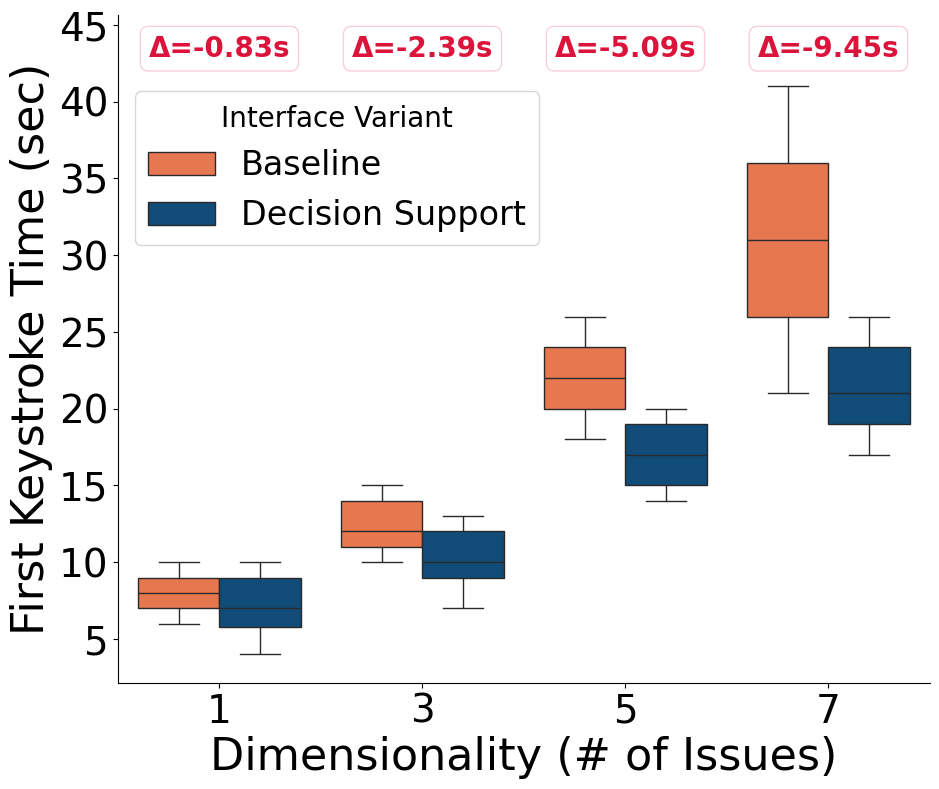}
          \caption{Average first keystroke time.}
          \label{fig:first-keystroke}
        \end{subfigure}        
        \caption{Primary objective outcomes as box plots: (a) In the Baseline condition, total human payoff declines sharply after three issues; Decision Support prevents this decline and maintains stable payoffs across dimensionalities (no statistically significant effect of dimensionality in Decision Support). (b) Total turns to agreement increase with dimensionality; Decision Support requires relatively fewer turns than Baseline at higher dimensionalities. (c) Average first keystroke time (excluding turn no. 1) increases with dimensionality in Baseline; Decision Support reduces this latency, especially at higher dimensionalities. $\Delta$ represents the difference between mean Decision Support and mean Baseline conditions ($\Delta = \text{Decision Support}_{\text{mean}} - \text{Baseline}_{\text{mean}}$).}
        \label{fig:payoff-turns-combined}        
        \Description{Three-panel figure of box plots: (a) total human payoff (vertical) across issue counts (horizontal) comparing Baseline and Decision Support, showing payoff decline in Baseline but preservation under Decision Support; (b) total turns (vertical) by issue count (horizontal), showing increased turns with complexity in Baseline and reduced turns under Decision Support; (c) average first keystroke time (vertical) by issue count (horizontal), showing increased latency with complexity in Baseline and reduced latency under Decision Support.}
      \end{figure*}

    \subsubsection{Total Human Payoff}
    \label{sec:human_payoff}

      To answer RQ1, we first examined the Baseline condition. A one-way repeated-measures ANOVA showed that human payoff significantly decreased as Dimensionality increased (\textit{F}(1.755, 54.407) = 28.36, \textit{p} $<$ .001, $\eta^2$ = 0.478), establishing the negative impact of complexity.
      To answer RQ2, a 2×4 repeated-measures ANOVA was conducted. We found a significant main effect of Interface and a significant Interface × Dimensionality interaction (summarized in Table~\ref{tab:summary_stats}), indicating that the Decision Support condition yielded higher payoffs and was most effective at preserving human payoff at higher Dimensionality levels (see Fig~\ref{fig:total-human-payoff}).
      Holm-adjusted post-hoc tests (Table~\ref{tab:posthoc_baseline}) showed that, in contrast to the Baseline, Decision Support left individual payoffs unchanged (n.s.), thereby preventing the decline seen in Baseline, while also simultaneously suppressing any significant payoff reallocation.

    \subsubsection{Total Turns to Agreement}
    \label{sec:negotiation_efficiency}
      In the Baseline condition, the number of turns required to reach an agreement increased significantly with Dimensionality (\textit{F}(3, 93) = 844.00, \textit{p} $<$ .001, $\eta^2$ = 0.965). This near-deterministic effect reflects that issue count directly determines negotiation length in our task structure, leaving little residual variance. We therefore treat it as a task-level manipulation check rather than evidence of a psychological process.
      The full 2×4 ANOVA showed a significant main effect of Interface and a significant interaction (see Table~\ref{tab:summary_stats} and Fig~\ref{fig:total-turns}), indicating that Decision Support reduced turns, particularly in higher-dimensional negotiations.
    \subsubsection{Total Chat Duration}
    \label{sec:chat_duration}
      There was a significant main effect of dimensionality on total chat duration in the Baseline condition, with Greenhouse-Geisser correction applied due to sphericity violation (Mauchly's W = 0.664, \textit{p} = .032), \textit{F}(2.317, 71.819) = 2615, \textit{p} $<$ .001, $\eta^2$ = 0.988.
      The 2×4 ANOVA likewise showed significant main effects of Interface and Dimensionality and a significant interaction (see Table~\ref{tab:summary_stats}), indicating the Decision Support interface substantially reduced chat duration as complexity increased.
    \subsubsection{Average First Keystroke Time}
    \label{sec:first_keystroke}
      We found a significant main effect of Dimensionality on average first keystroke time in the Baseline condition (\textit{F}(3, 93) = 2611, \textit{p} $<$ .001, $\eta^2$ = 0.988), indicating first-keystroke latency increased sharply with task dimensionality.
      The 2×4 ANOVA too showed significant main effects of Interface and Dimensionality and a significant interaction (see Table~\ref{tab:summary_stats} and Fig~\ref{fig:first-keystroke}).

  \subsection{Process \& Behavioral Metrics}
  \label{sec:process_metrics}
    \subsubsection{Backtracking Frequency}
    \label{sec:backtracking}
      In the Baseline condition, backtracking frequency rose sharply with Dimensionality (\textit{F}(3, 93) = 549.00, \textit{p} $<$ .001, $\eta^2$ = 0.947).
      The Decision Support tool significantly mitigated this. The 2×4 ANOVA showed a significant main effect of Interface and an Interface × Dimensionality interaction (see Table~\ref{tab:summary_stats}), demonstrating the tool's effectiveness in helping users maintain a more consistent negotiation path.
    \subsubsection{Concession Behavior}
    \label{sec:concession_behavior}
      In the Baseline condition, the average concession magnitude was significantly affected by Dimensionality (\textit{F}(1.903, 58.994) = 145.20, \textit{p} $<$ .001, $\eta^2$ = 0.824), as was the total number of concessions made (\textit{F}(3, 93) = 515.80, \textit{p} $<$ .001, $\eta^2$ = 0.943).
      The 2×4 ANOVAs showed significant main effects of Interface and significant interactions for both average concession magnitude and number of concessions (see Table~\ref{tab:summary_stats}), indicating the tool altered concession strategies.
    \subsubsection{Sequence Entropy}
    \label{sec:sequence_entropy}
      In the Baseline condition, the entropy of negotiation sequences was significantly affected by Dimensionality (\textit{F}(1.679, 52.039) = 17.54, \textit{p} $<$ .001, $\eta^2$ = 0.361), indicating less predictable trajectories in more complex tasks.
      The Decision Support tool significantly reduced this unpredictability.
      The 2×4 ANOVA showed a significant main effect of Interface and a significant interaction (see Table~\ref{tab:summary_stats}), suggesting the tool helped structure the negotiation process more effectively.

    \subsubsection{Joint Payoff}
    \label{sec:joint_payoff}
      A 2×4 repeated-measures ANOVA of the joint human-AI payoff revealed a significant main effect of Interface (\textit{F}(1, 31) = 13.27, \textit{p} $<$ .001, $\eta_p^2$ = 0.300), with the Decision Support condition producing higher joint payoffs overall. The main effect of Dimensionality and the Interface × Dimensionality interaction were not significant (\textit{p}'s $>$ .05).

    \begin{table*}[h!]
      \centering
      \caption{Summary of 2×4 Repeated-Measures ANOVA Results for Key Dependent Variables. All \textit{p} $<$ .001 unless otherwise specified. Fractional dfs reflect Greenhouse–Geisser correction for sphericity violations}
      \label{tab:summary_stats}
      \resizebox{\textwidth}{!}{%
      \begin{tabular}{l l l l}
      \toprule
      \textbf{Dependent Variable} & \textbf{Main Effect of Interface} & \textbf{Main Effect of Dimensionality} & \textbf{Interface × Dimensionality Interaction} \\
      \midrule
      \multicolumn{4}{l}{\textit{Objective Outcomes}} \\
      \quad Total Human Payoff & \textit{F}(1, 31) = 59.87, $\eta_p^2$ = 0.659 & \textit{F}(2.318, 71.854) = 28.27, $\eta_p^2$ = 0.477 & \textit{F}(1.860, 57.675) = 11.48, $\eta_p^2$ = 0.270 \\
      \quad Total Turns & \textit{F}(1, 31) = 237.77, $\eta_p^2$ = 0.885 & \textit{F}(3, 93) = 879.22, $\eta_p^2$ = 0.966 & \textit{F}(2.329, 72.206) = 26.92, $\eta_p^2$ = 0.465 \\
      \quad Total Chat Duration & \textit{F}(1, 31) = 730.50, $\eta_p^2$ = 0.959 & \textit{F}(2.212, 68.557) = 7849.90, $\eta_p^2$ = 0.996 & \textit{F}(2.288, 70.917) = 227.50, $\eta_p^2$ = 0.880 \\
      \quad Avg. First Keystroke Time & \textit{F}(1, 31) = 1085.3, $\eta_p^2$ = 0.972 & \textit{F}(1.824, 56.558) = 3269.4, $\eta_p^2$ = 0.991 & \textit{F}(2.132, 66.088) = 399.3, $\eta_p^2$ = 0.928 \\      
        \midrule
        \multicolumn{4}{l}{\textit{Process \& Behavioral Metrics}} \\
      \quad Backtracking Frequency & \textit{F}(1, 31) = 75.11, $\eta_p^2$ = 0.708 & \textit{F}(2.456, 76.148) = 702.91, $\eta_p^2$ = 0.958 & \textit{F}(3, 93) = 20.89, $\eta_p^2$ = 0.403 \\
      \quad Avg. Concession Mag. & \textit{F}(1, 31) = 90.63, $\eta_p^2$ = 0.745 & \textit{F}(1.950, 60.446) = 264.16, $\eta_p^2$ = 0.895 & \textit{F}(3, 93) = 20.04, $\eta_p^2$ = 0.393 \\
      \quad Num. Concessions & \textit{F}(1, 31) = 23.40, $\eta_p^2$ = 0.430 & \textit{F}(3, 93) = 1050.32, $\eta_p^2$ = 0.971 & \textit{F}(3, 93) = 11.03, $\eta_p^2$ = 0.262 \\
      \quad Sequence Entropy & \textit{F}(1, 31) = 124.87, $\eta_p^2$ = 0.801 & \textit{F}(2.161, 66.989) = 10.72, $\eta_p^2$ = 0.257 & \textit{F}(1.8, 66.989) = 7.40, \textit{p} = .002, $\eta_p^2$ = 0.193 \\
        \midrule
        \multicolumn{4}{l}{\textit{Subjective Experience}} \\
      \quad TLX Mental Demand & \textit{F}(1, 31) = 58.69, $\eta_p^2$ = 0.654 & \textit{F}(3, 93) = 1185.91, $\eta_p^2$ = 0.975 & \textit{F}(2.320, 71.909) = 16.69, $\eta_p^2$ = 0.350 \\
      \quad TLX Temporal Demand & \textit{F}(1, 31) = 63.56, $\eta_p^2$ = 0.672 & \textit{F}(3, 93) = 739.50, $\eta_p^2$ = 0.960 & \textit{F}(3, 93) = 23.85, $\eta_p^2$ = 0.435 \\
      \quad TLX Effort & \textit{F}(1, 31) = 174.22, $\eta_p^2$ = 0.849 & \textit{F}(3, 93) = 633.45, $\eta_p^2$ = 0.953 & \textit{F}(3, 93) = 32.45, $\eta_p^2$ = 0.511 \\
      \quad TLX Frustration & \textit{F}(1, 31) = 132.54, $\eta_p^2$ = 0.810 & \textit{F}(3, 93)) = 925.40, $\eta_p^2$ = 0.968 & \textit{F}(3, 93) = 36.09, $\eta_p^2$ = 0.538 \\
      \quad TLX Performance & \textit{F}(1, 31) = 1.923, \textit{p} = .175, $\eta_p^2$ = 0.058 & \textit{F}(3, 93) = 822.135, $\eta_p^2$ = 0.964 & \textit{F}(1.942, 60.201) = 3.701, \textit{p} = .032, $\eta_p^2$ = 0.107 \\             
      \quad Confidence & \textit{F}(1, 31) = 63.56, $\eta_p^2$ = 0.672 & \textit{F}(3, 93) = 739.50, $\eta_p^2$ = 0.960 & \textit{F}(3, 93) = 23.85, $\eta_p^2$ = 0.435 \\
      \quad Strategy Satisfaction & \textit{F}(1, 31) = 73.99, $\eta_p^2$ = 0.705 & \textit{F}(3, 93) = 470.23, $\eta_p^2$ = 0.938 & \textit{F}(2.204, 68.325) = 97.68, $\eta_p^2$ = 0.759 \\
      \bottomrule
      \end{tabular}%
      }
    \end{table*}

  \subsection{Subjective Experience Outcomes}
  \label{sec:subjective_outcomes}
    To assess subjective experience, we analyzed self-reported cognitive load using the NASA-TLX subscales, as well as self-reported confidence and strategy satisfaction.
    \subsubsection{Cognitive Load (NASA-TLX)}
    \label{sec:cognitive_load}
      In the Baseline condition, cognitive load increased significantly with Dimensionality across multiple subscales, including Mental Demand (\textit{F}(3, 93) = 744.50, \textit{p} $<$ .001, $\eta^2$ = 0.960), Temporal Demand (\textit{F}(3, 93) = 642.10, \textit{p} $<$ .001, $\eta^2$ = 0.954), Effort (\textit{F}(3, 93) = 589.70, \textit{p} $<$ .001, $\eta^2$ = 0.950), Frustration (\textit{F}(2.121, 65.740) = 890.80, \textit{p} $<$ .001, $\eta^2$ = 0.966) and Performance (\textit{F}(3, 93) = 483.50, \textit{p} $<$ .001, $\eta^2$ = 0.940). As expected, Physical Demand was not affected. 
      The Decision Support tool significantly mitigated cognitive load. 
      The 2×4 ANOVA showed significant main effects of Interface and significant Interface × Dimensionality interactions for Mental Demand, Performance, Temporal Demand, and Effort (see Table~\ref{tab:summary_stats}), confirming the tool was most effective at reducing cognitive burden in high-dimensionality tasks.
    \subsubsection{Confidence and Strategy Satisfaction}
    \label{sec:confidence_satisfaction}
      In the Baseline condition, participants' confidence (\textit{F}(2.153, 66.734) = 272.00, \textit{p} $<$ .001, $\eta^2$ = 0.898) and satisfaction with their strategy (\textit{F}(2.608, 80.848) = 557.90, \textit{p} $<$ .001, $\eta^2$ = 0.947) declined significantly as Dimensionality increased.
      The Decision Support tool substantially improved both measures.
      The 2×4 ANOVA showed significant main effects of Interface and significant interactions for both confidence and strategy satisfaction (see Table~\ref{tab:summary_stats}), indicating that users felt more confident and satisfied when using the tool in the most complex negotiations.

      Finally, the SUS results: The interface achieved an average SUS score of 81.02, indicating excellent usability.

  \subsection{Summary of Findings}
  \label{sec:summary_findings}

    In response to RQ1, our results consistently show that increasing negotiation Dimensionality in the Baseline condition significantly harmed objective outcomes (lower payoff), process metrics (more turns, more backtracking), and subjective experience (lower satisfaction, higher frustration).
    In response to RQ2, the Decision Support interface successfully and significantly mitigated these negative effects across all categories of metrics. Significant Interface × Dimensionality interactions on key variables confirm that the tool's benefits were most pronounced in the more complex 5- and 7-issue negotiations.

\section{DISCUSSION}
\label{sec:discussion}

    We interpret findings relative to prior work on cognitive asymmetry and visualization for decision-making, then discuss theory, design, limitations, and future directions.

        \begin{figure}[!htb]
            \centering
            \begin{subfigure}[b]{0.8\linewidth}
                \centering
                \includegraphics[width=\linewidth]{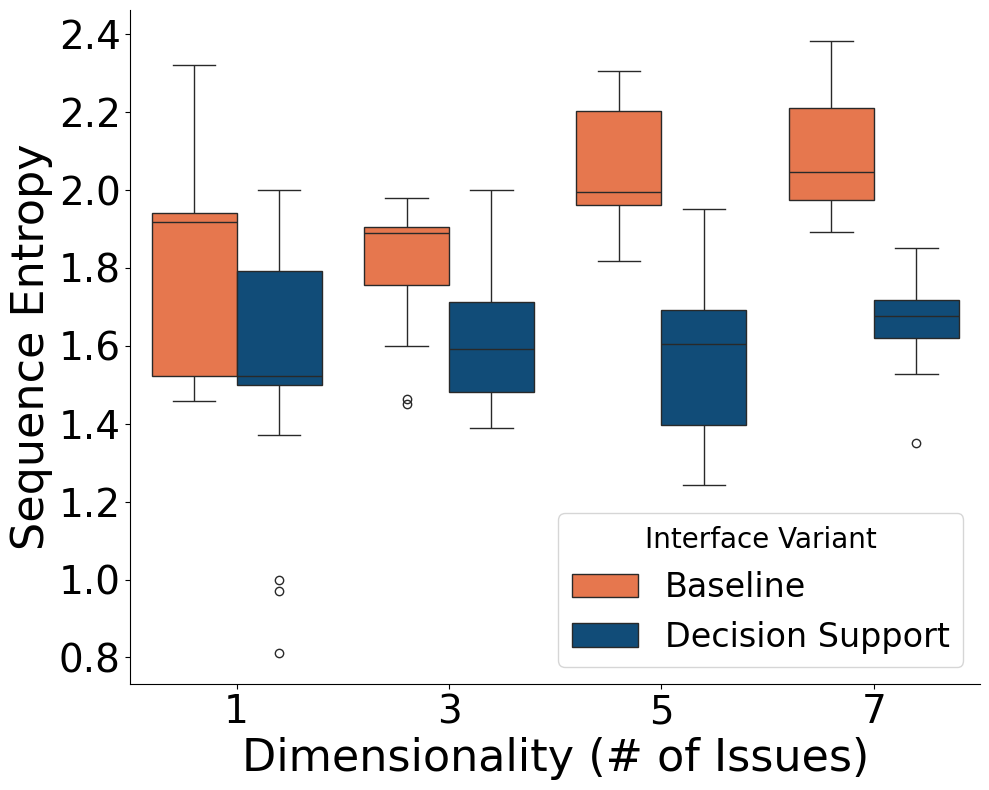}
                \caption{Sequence entropy scaling.}
                \label{fig:sequence-entropy}
            \end{subfigure}

            
            \begin{subfigure}[b]{0.80\linewidth}
                \centering
                \includegraphics[width=\linewidth]{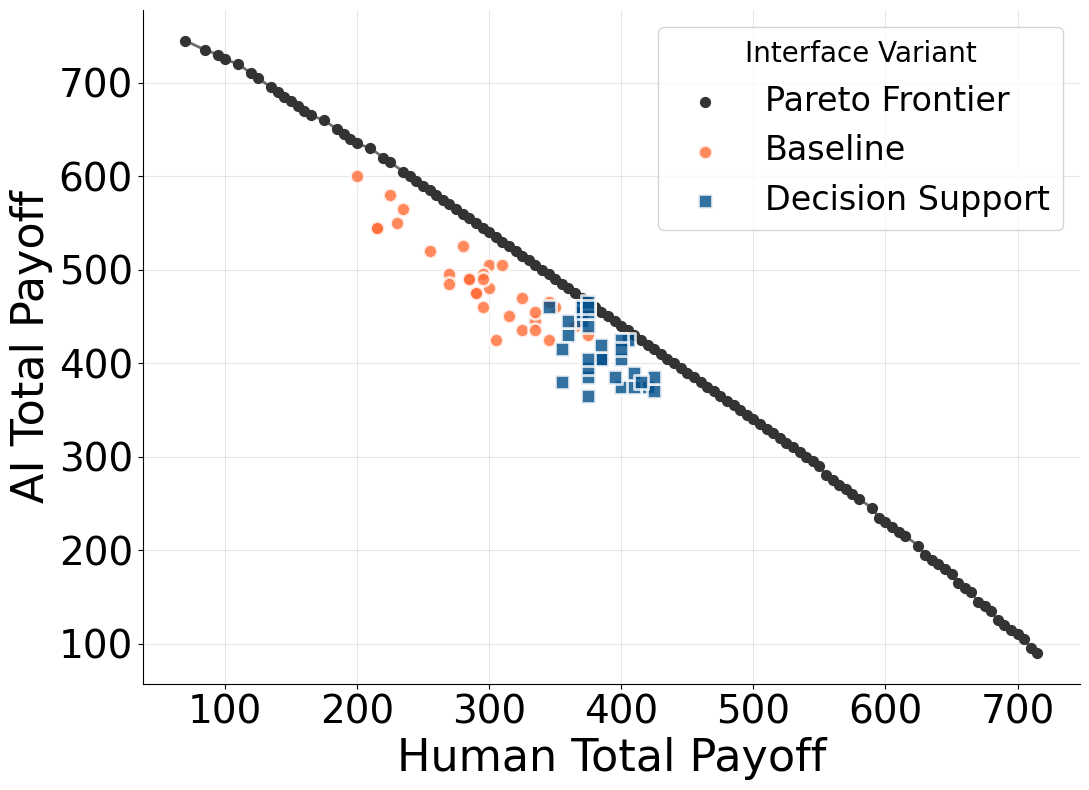}
                \caption{Pareto proximity (7 issues).}
                \label{fig:pareto-7}
            \end{subfigure}
            \caption{Joint value optimization and exploration structure: (a) Box plot of sequence entropy across dimensionalities, showing greater uncertainty at higher dimensions in the Baseline condition. (b) Scatterplot of total human + AI payoffs for 7 issues, with Decision Support points clustered nearer the center and closer to the Pareto frontier.}
            \label{fig:pareto-entropy-combined}
            \Description{Two-panel figure combining joint-value and exploration structure: (a) a box plot of sequence entropy across dimensionalities comparing conditions, where Baseline shows a steep rise and Decision Support flattens that increase. Panels encode condition with color and illustrate that DS advantage increases with dimensionality; (b) scatter plot of negotiated joint payoffs versus theoretical Pareto frontier for 7-issue sessions, showing Decision Support sessions clustered closer to the frontier with more balanced outcomes compared with Baseline}
        \end{figure}

    \subsection{Impact of Dimensionality on Human-AI Negotiation (RQ1)}
    \label{sec:impact_dimensionality}

        Our findings empirically confirm that increasing issue dimensionality in human-AI negotiation degrades human performance. We establish this based on objective measures, including decreases in human payoff, reduced negotiation efficiency (Figure~\ref{fig:total-turns}), increased deliberation time (Figure~\ref{fig:first-keystroke}), and subjective indicators.

        Our findings reveal a selective \textbf{plateau–cliff effect}: human payoff and sequence entropy degrade sharply after three issues (Figures~\ref{fig:total-human-payoff},~\ref{fig:sequence-entropy}), while other metrics decline gradually. In Baseline, payoff remains stable from one to three issues but drops significantly at five and seven (Table~\ref{tab:posthoc_baseline}). Entropy mirrors this pattern, increasing sharply at five issues before plateauing. Subjective burden and concession pacing show smoother changes, indicating the cliff is cognitive rather than motivational.
        This provides evidence for a \textbf{bounded integrative window}: up to three issues, participants manage trade-offs; beyond that, combinatorial load causes breakdown in sequence control, longer decision latencies (Figure~\ref{fig:first-keystroke}), and lower payoff. The smooth rise in frustration suggests compensatory heuristics (larger concessions) rather than fatigue, aligning with human-human negotiation research \cite{warsitzka2024expanding}.

        The plateau-cliff pattern identifies precise targets for visualization support. Increased dimensionality creates friction through decision latency and inefficient exploration, exactly where integrative capacity reaches its threshold. While derived from integrative, equally-weighted issues in property rental, this threshold represents a key design insight for multi-issue tasks. We hypothesize similar collapses occur when complexity exceeds human tracking bandwidth in other domains lacking cross-dimensional support, underscoring the need for adaptive support.

    \subsection{Effectiveness of the Decision Support Condition (RQ2)}
    \label{sec:effectiveness_ds_widget}
        Our findings show that the decision support interface successfully mitigated the performance collapse observed at high dimensionalities. 
        We conceptualize its role as a \textbf{cognitive prosthesis}: an extension of the user's own mind that offloads the mechanical burdens of tracking complexity, thereby preserving their strategic agency. 
        Like a physical prosthesis that restores mobility while the user controls direction, our tool removed coordination friction without prescribing choices. It thus stabilized payoffs and reduced subjective load precisely where unaided performance degraded.

        \subsubsection{Decision Support as Cognitive Prosthesis: Key Findings}
        \label{sec:prosthetic_function}
    
            The 2×4 ANOVA results (Table~\ref{tab:summary_stats}) confirm this prosthetic role through four complementary functions. First, the interface preserved human payoff as issue dimensionality increased (Figure~\ref{fig:total-human-payoff}); planned post-hoc contrasts (1--3, 3--5, 5--7 issues) showed no significant differences within the Decision Support condition, confirming the tool prevented performance erosion. Second, negotiations became more efficient: fewer turns to agreement (Figure~\ref{fig:total-turns}) and reduced first-keystroke latency (Figure~\ref{fig:first-keystroke}) emerged while participants retained full control over proposals and concessions.

            Third, sequence entropy, a measure of proposal variability, declined under Decision Support (Figure~\ref{fig:sequence-entropy}), reflecting more focused search and steadier concession dynamics. Fourth, cognitive load measures including Mental Demand, Temporal Demand, and Effort (Table~\ref{tab:summary_stats}) decreased and participants reported higher confidence, particularly at higher dimensionalities. Together, these effects indicate the tool acted as a \textit{cognitive prosthesis}, offloading integrative burdens while preserving user agency. These benefits scaled with task complexity; we next examine differential effectiveness and underlying mechanisms.

        \subsubsection{How Tool Effectiveness Scales with Issue Count}
        \label{sec:scaling_effects}
            \begin{figure*}[h]
                    \centering
                    \includegraphics[width=0.9\linewidth]{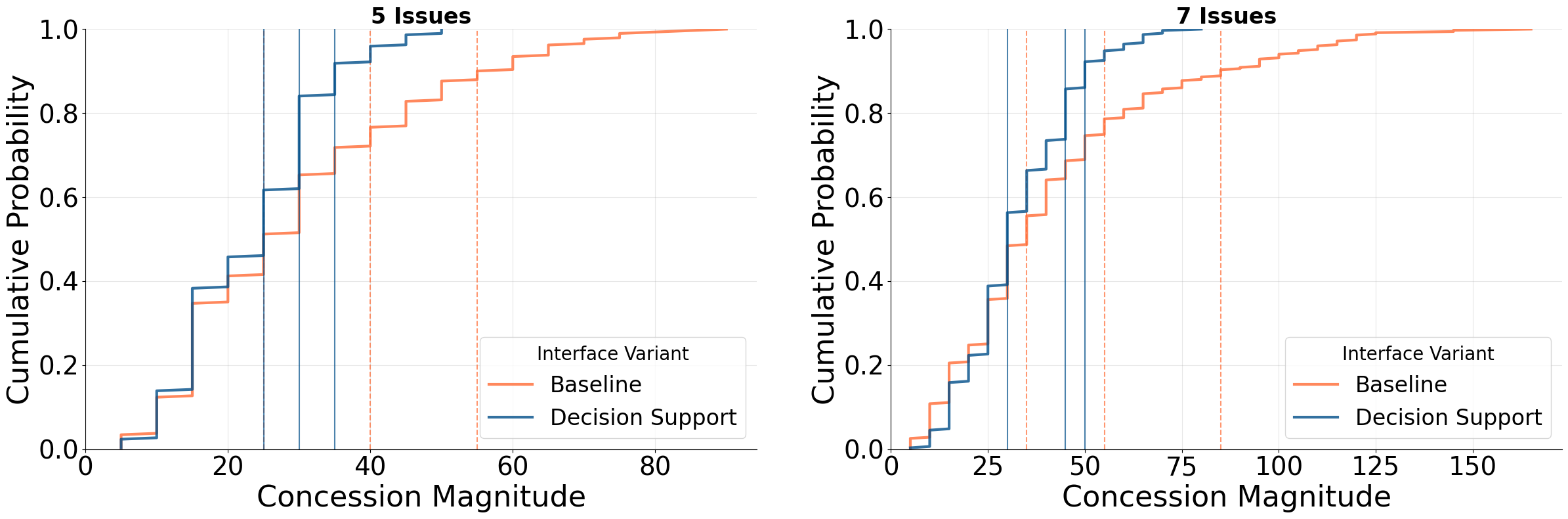}
                    \caption{Concession magnitude CDF (5 \& 7 issues): CDFs of concession magnitudes by condition, showing that Decision Support shifts mass toward smaller, more frequent concessions (steeper rise at lower magnitudes), especially at higher dimensionality, indicating pacing recalibration and fewer large overshoot–correct cycles.}
                    \label{fig:concession-magnitude-CDF}
                    \Description{CDF plots of concession magnitudes for 5- and 7-issue conditions, split by Baseline and Decision Support. The curves show that Decision Support shifts mass toward smaller, more frequent concessions (steeper rise at lower magnitudes), especially at higher dimensionality, indicating pacing recalibration and fewer large overshoot-correct cycles. Axes: concession magnitude (horizontal) and cumulative probability (vertical).}
            \end{figure*}

            \begin{figure*}[h]
                \centering
                \includegraphics[width=0.75\linewidth]{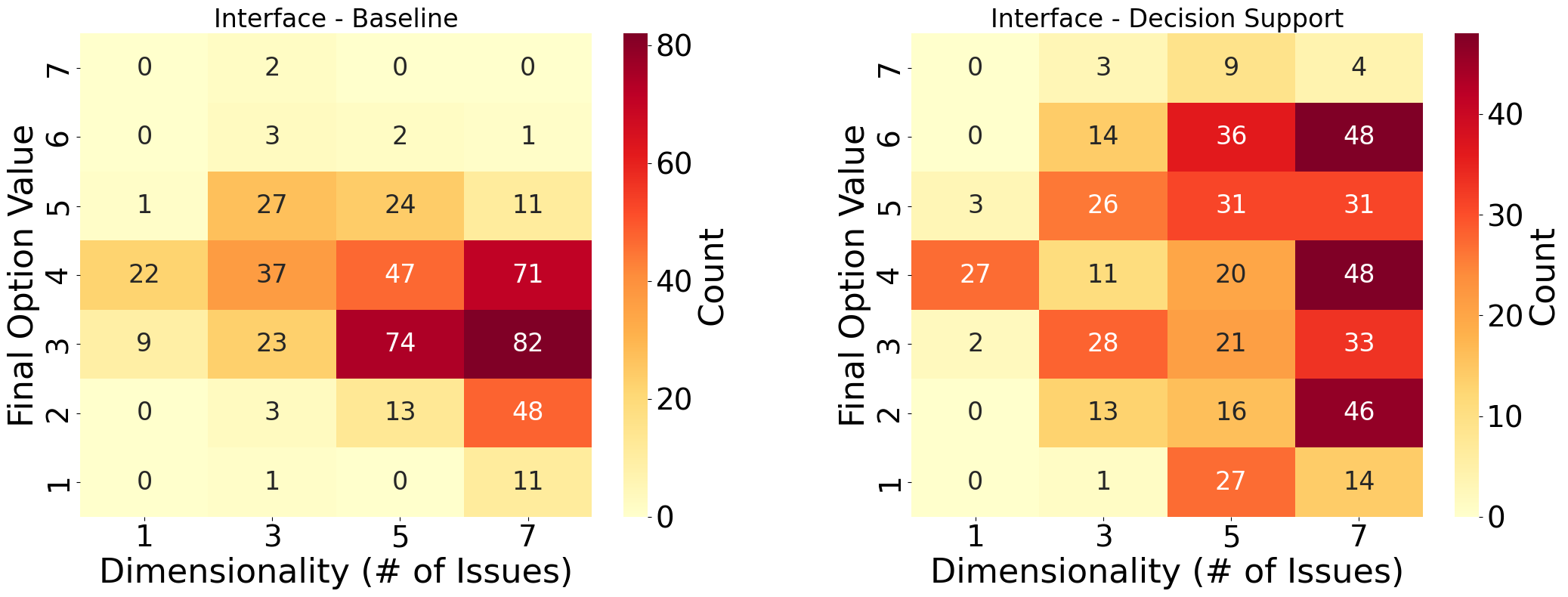}
                \caption{Heatmap of final option selections (2‑D histogram of counts) for 7 issues): In Baseline, negotiators concentrate on mid–low option values, whereas in Decision Support, participants exhibit a broader, more even spread across intermediate options while still avoiding extreme high-risk choices.}
                \label{fig:final-options-distribution}
                \Description{Heatmap of final option selections aggregated across the seven issues. The distribution highlights behavioral differences: Baseline negotiators concentrate on mid-low option values, while Decision Support users show a broader, more even spread across intermediate options and avoid extreme high-risk choices. The plot's horizontal axis lists option values and vertical axis shows frequency of selection.}
            \end{figure*}

            Evidence converges that Decision Support yields its largest benefits at higher dimensionalities (5–7 issues), across outcome efficiency and process recalibration.
            Decision Support substantially reduced the distance to Pareto-optimal agreements, with benefits scaling proportionally to dimensionality. At seven issues, dyads using the tool negotiated agreements 12.4 payoff units closer to the Pareto frontier compared to Baseline (Figure~\ref{fig:pareto-7}). This advantage declined systematically as dimensionality decreased: 8.94 units at five issues, 5.50 units at three issues, and only 0.09 units at one issue. This gradient confirms the tool addresses combinatorial complexity. The scatterplot (Figure~\ref{fig:pareto-7}) further reveals that Decision Support improved proximity and concentrated outcomes near the center, indicating more balanced agreements.

            \paragraph{Process recalibration}
                Two behavioral shifts drive these outcome improvements:

                \textit{(a) Finer concession pacing.}
                Under Decision Support, concession pacing became more controlled and fine-grained. Smaller, more frequent adjustments replaced large overshoot-correct cycles. The concession-magnitude CDFs for 5- and 7-issue sessions (Figure~\ref{fig:concession-magnitude-CDF}) show a shift toward smaller, more frequent adjustments.

\begin{figure*}[h]
                    \centering
                    \includegraphics[width=0.8\linewidth]{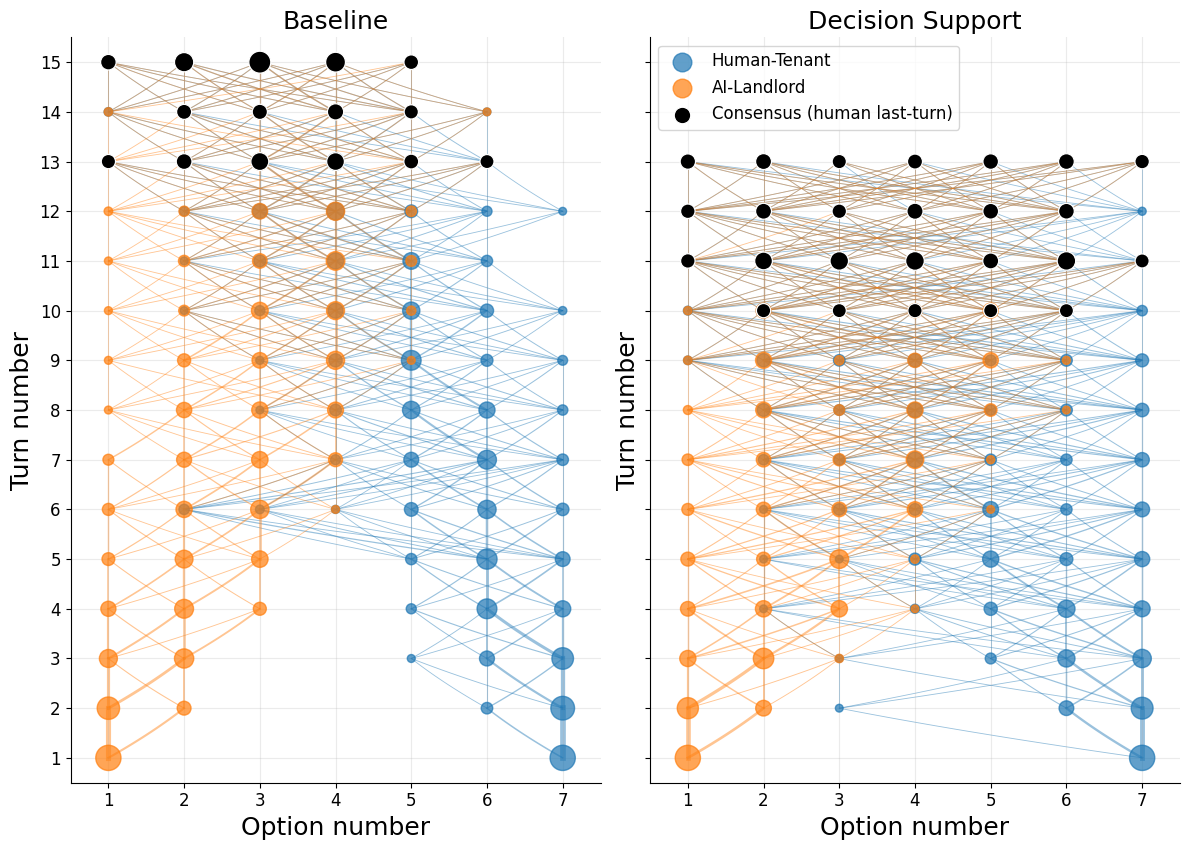}
                    \caption{Option transition network at high dimensionality (7 issues): Baseline sessions show sparse early exploration (turns $\sim$2--6) with convergence occurring later (turns $\sim$12--15), whereas Decision Support sessions exhibit earlier exploration and faster convergence (turns $\sim$10--13).}
                    \label{fig:transition-network-plot}
                    \Description{Network diagram of option-to-option transitions over negotiation turns (7-issue sessions). Nodes represent option states, edges represent transitions between states across turns, and edge thickness encodes transition frequency. Decision Support networks show denser, earlier convergence to higher-value regions (greater transition density in turns ~10--13) whereas Baseline networks remain more diffuse and extend convergence later (toward turn 15), illustrating trajectory efficiency differences.}
                \end{figure*}

                \textit{(b) Improved search topology.}  
                Decision Support restructured how participants explored the solution space. The distribution of final option selections (Figure~\ref{fig:final-options-distribution}) indicates participants made broader, more uniform intermediate choices (options 2--5), while still avoiding extreme high-risk choices (options 6--7). Network visualizations of option-to-option transitions (Figure~\ref{fig:transition-network-plot}) at high dimensionality (7 issues) show Baseline sessions exhibited sparse early exploration with late convergence, while Decision Support sessions demonstrated denser early exploration and faster convergence to high-value regions.

                These improvements in final outcomes and negotiation processes demonstrate Decision Support is most effective at higher dimensions, precisely where unaided human performance collapses. We next explain how the design offloads this complexity.

        \subsubsection{Cognitive Offloading Mechanisms}
        \label{sec:cognitive-offloading-mechanisms}

\begin{figure*}[h]
                \centering
                \begin{subfigure}[b]{0.44\linewidth}
                    \centering
                    \includegraphics[width=\linewidth]{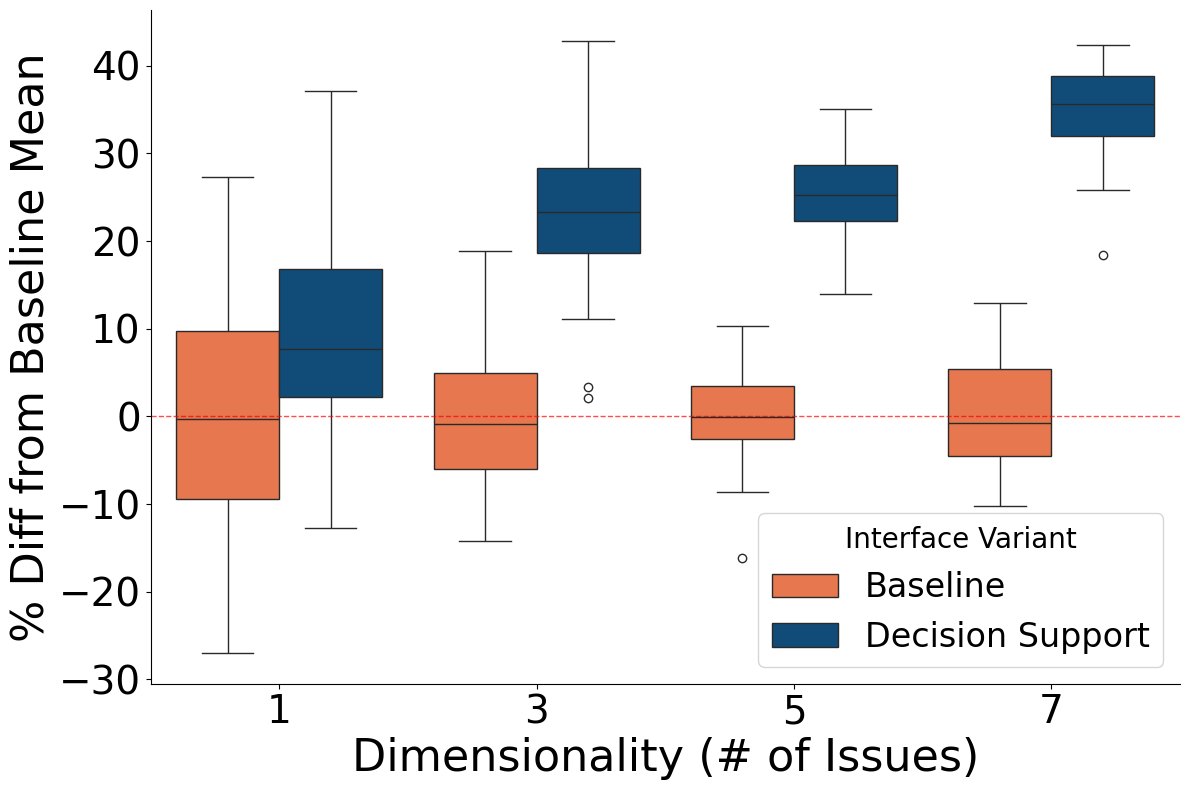}
                    \caption{Total chat duration (normalized).}
                    \label{fig:duration-normalized}
                \end{subfigure}\hfill
                \begin{subfigure}[b]{0.44\linewidth}
                    \centering
                    \includegraphics[width=\linewidth]{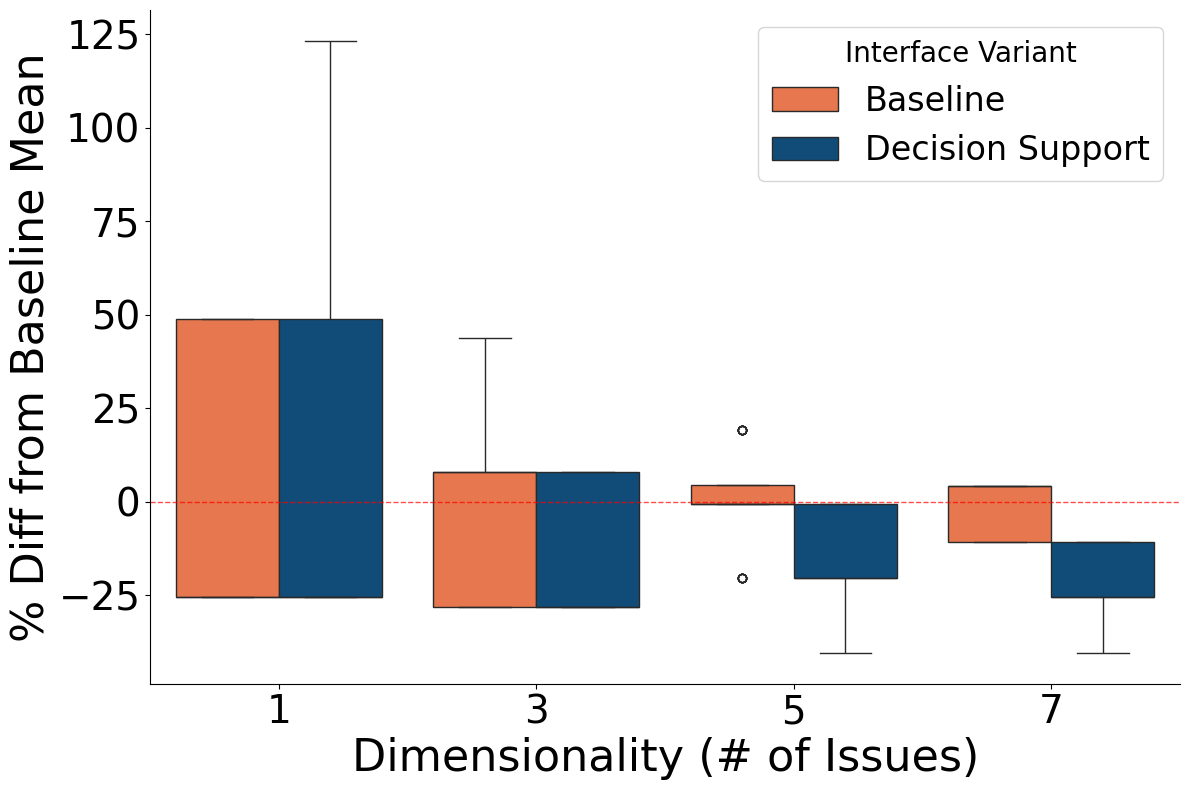}
                    \caption{TLX Temporal Demand.}
                    \label{fig:tlx-temporal-demand}
                \end{subfigure}
                \caption{Objective vs. Subjective time, normalized to baseline condition: Decision Support yields longer interactions relative to Baseline, but lower perceived temporal demand at high dimensionality, indicating a reframed temporal experience.}
                \label{fig:duration-temporal-demand}
                \Description{Two-panel figure: (a) normalized total chat duration by condition and dimensionality, showing Decision Support sessions longer on average; (b) TLX Temporal Demand by condition and dimensionality, showing lower perceived temporal demand under Decision Support despite longer durations.}
            \end{figure*}

            The payoff preservation, efficiency gains, and reduced subjective burden described above stem from \textbf{cognitive offloading}: the interface transforms difficult memory tasks into easier perceptual tasks through behavioral, perceptual, and computational mechanisms:

            \paragraph{Behavioral: Structured exploration and temporal recalibration.}  
            Decision Support prevented the sharp entropy increases observed in Baseline as dimensionality grew (Figure~\ref{fig:sequence-entropy}), stabilizing exploration across problem complexity. The tool reduced first-keystroke latency (Figure~\ref{fig:first-keystroke}) while extending total chat duration (Figure~\ref{fig:duration-normalized}), yet participants reported significantly lower TLX Temporal Demand (Figure~\ref{fig:tlx-temporal-demand}). This dissociation between objective duration (longer) and subjective temporal demand (lower) suggests an \textbf{expectations-elevation effect}: making trade-offs visible and highlighting proximity to optima may recalibrate how users perceive time pressure, allowing more thorough engagement despite extended interaction. A similar pattern emerged in performance self-assessments---participants rated their TLX performance lower on average in the Decision Support condition despite improved objective payoffs and reduced entropy, though this difference was not statistically significant (Table~\ref{tab:summary_stats}, $p=0.175$). Together, these patterns suggest that decision support may elevate user expectations about what constitutes ``good'' performance and efficient use of time. This interpretation is based on single TLX items and should be treated as provisional; future work should employ multi-item scales to better understand how decision support affects self-assessment and expectations.

            \paragraph{Perceptual: Decomposing complexity through dual-channel visualization.}
            The \textit{Negotiation Horizon Grid} converts memory retrieval into visual search, while the \textit{Global Convergence Panel} summarizes progress without mental tabulation. This separation addresses distinct cognitive demands: spatial tracking (which options are acceptable) versus temporal monitoring (how close are we to agreement), preventing overload while maintaining a complete picture.

            \paragraph{Computational: Bayesian filtering for agency preservation.}
            Underpinning these visuals, Bayesian inference filters communication noise, visualizing \textit{uncertainty} rather than dictating \textit{moves}. This extends established cognitive science principles \cite{kersten2017heuristics, kim2019bayesian, prabhudesai2023understanding} where external representations reduce load by making structure explicit and transforming memory into perception. Our contribution applies this framework to adversarial settings with hidden preferences, providing normative guidance on where uncertainty remains while preserving user control.

        Having shown where humans break down and how decision-support tools help, we now derive actionable design principles and generalize these into theory.

    \subsection{Design Implications for Human-AI Negotiation Support}
    \label{sec:design_implications}
        We distill four actionable design principles:

        \begin{enumerate}
            \item \textit{Adaptive prosthesis:} Employ minimal support for simpler negotiations; trigger richer cues only after detecting the integrative threshold (here, beyond three issues) to prevent selective collapse without over-structuring simple cases.  

            \item \textit{Dual-channel offloading:} Separate spatial tracking (which options are acceptable) from temporal monitoring (how close are we to agreement). Our Negotiation Horizon Grid handles the former; the Global Convergence Panel handles the latter. This separation prevents overload while maintaining a complete picture.

            \item \textit{Prosthetic, not prescriptive stance:} Reduce friction while preserving outcomes. For instance, highlight that ``both parties keep proposing low-value options'' rather than suggesting ``offer \$1200 rent''. Our stable payoffs (Figure~\ref{fig:total-human-payoff}) and high usability (SUS=81.02) confirm this approach has low adoption cost and preserves agency.

            \item \textit{Fairness through waste removal:} Focus on removing inefficiencies (e.g., showing which proposals were already rejected) rather than redistributing gains. Users accept such assistance as fair support tools, not biased intervention.
        \end{enumerate}

    These principles, validated in our controlled property rental negotiation study, offer promising directions for other contexts. Potential applications include multi-criteria product recommenders (reducing choice overload), medical decision aids (structuring uncertainty about treatment options), and financial planning tools (progress feedback without prescribing investments). The key insight: support becomes acceptable when it removes structural bottlenecks without steering final allocations.

    \subsection{Theoretical Contributions}
    \label{sec:theoretical_contributions}

        Our findings advance theory on human-AI negotiation support through three empirically grounded mechanisms and one tentative observation. Together, these demonstrate a principle we term \textbf{Cognitive Harmony}: augmenting human cognition to reduce inefficiency while preserving fairness and agency; below we state each mechanism succinctly and note limits on generalization.

        \paragraph{Bounded Integrative Processing.}
        We establish a discrete cognitive limit: the \textit{plateau-cliff effect} where performance collapses sharply beyond three issues (Figures~\ref{fig:total-human-payoff}, ~\ref{fig:sequence-entropy}).        
        This extends cognitive load theory by identifying a domain-specific threshold for adversarial coordination, clarifying precisely when support becomes necessary. Unlike prior work that assumes linear degradation, our data reveal a non-linear boundary that informs adaptive system design. 
        Caveat: this boundary is empirical and task-dependent; different issue types, repeated interactions, or heterogeneous weights may shift or remove the threshold.

        \paragraph{Cognitive Prosthesis.} 
        External visualizations that separate spatial and temporal information (our dual-channel design) can preserve performance at higher dimensionalities without prescribing moves. In short, the interface extended users' capacity for integrative tracking while leaving strategic choice to participants. Applicability: demonstrated here for integrative property rental negotiation setup; generalization to zero-sum or categorical issues requires validation.

        \paragraph{Fairness Through Procedural Neutrality.}
        We propose \textit{procedural neutrality} as a practical criterion: supports that remove inefficiencies (e.g., redundant proposals, uncertainty) without reallocating gains are more likely to be perceived as fair. Empirically, this is consistent with stable payoff distributions and high usability in our study. Limitations: fairness perceptions may change when tools make prescriptive recommendations or intervene in strategic allocation.

        \textit{Tentative observation:} We report an \textit{expectations-elevation} pattern: richer visibility of optima coincided with lower self-rated performance despite objective gains. This is based on a single TLX item and should be treated as provisional; more targeted measures are needed to confirm the effect.

        \paragraph{Integration: The Cognitive Harmony Principle.}
        These mechanisms collectively instantiate the principle of \textit{Cognitive Harmony}.
        We offer a four-step prosthetic design cycle as a heuristic:
        (1) detect the cognitive cliff, (2) provide targeted support, (3) monitor expectation shifts, and (4) enforce procedural neutrality.
        This principle, derived from our case study of a rental negotiation scenario, generalizes beyond negotiation to any high-dimensional adversarial human-AI coordination task where agency preservation is paramount.

\section{LIMITATIONS}
\label{sec:limitations}
    Our findings should be interpreted in light of several design choices made to prioritize internal validity and isolate cognitive mechanisms. These choices, while necessary for experimental control, define clear boundaries for the generalizability of our results.

    First, our study employed a highly controlled task environment that simplified the complexities of real-world negotiation. We focused on cognitive factors by modeling the interaction as a combinatorial task with purely integrative issues, fixed roles, and equally weighted preferences. This approach intentionally minimized the influence of social dynamics (e.g., trust, rapport) and structural complexities (e.g., mixed-motive issues, power shifts) to isolate the effect of dimensionality on user performance. Consequently, our findings speak more to cognitive load than to the full spectrum of negotiation behavior.

    Second, the experimental design had specific methodological constraints. We used coarse steps in dimensionality (1, 3, 5, 7 issues) and a convenience sample of 32 participants. These choices were pragmatic trade-offs to manage participant fatigue within a single session and ensure sufficient statistical power for detecting large main effects, rather than smaller interaction effects or individual differences. This leaves the precise location of the performance ``cliff'' and its generalizability to a broader population as open questions.

    Third, we used a single AI agent configuration: a GPT-4 model with a rational, utility-maximizing persona. This consistency was crucial for reducing strategic variance and attributing user performance differences to our experimental manipulations. However, it means our results do not account for how users might interact with agents employing different strategies, such as prosocial, deceptive, or adaptive behaviors.

    Fourth, our within-subjects design did not fully counterbalance interface order; the baseline condition always preceded the decision support tool. We chose this fixed order to prevent tool exposure from influencing baseline performance and to observe unassisted capacity limits before augmentation. However, this design introduces the possibility of order effects, such as learning or fatigue, which could influence subjective ratings and performance in the latter condition.

\section{FUTURE WORK}
\label{sec:future_work}
    Building on the observed plateau–cliff effect and the mitigating role of decision support, we outline four directions for extending this work.  

    \paragraph{Richer negotiation structures.} Future studies should introduce heterogeneous weights, mixed integrative–distributive issues, and explicit complementarities to test whether the inflection point shifts. Exploring finer granularity between three and five issues, as well as hyper-complex (10–14+) spaces, can help parameterize when assistance should escalate from passive visualization to proactive suggestions.

    \paragraph{Strategic diversity and interface transparency.} Expand beyond rational agents to include prosocial, deceptive, or adaptive personas. Interfaces could incorporate strategic cues, volatility diagnostics, and adjustable transparency modes. Compare heuristic versus Bayesian elicitation to clarify when lightweight approximations suffice.  

    \paragraph{Longitudinal and ecological methods.} Multi-session studies and in-situ deployments can distinguish novelty from durable learning and reveal workplace constraints. Shared trace corpora with standardized annotations would enable reproducible benchmarking.  

    \paragraph{Personalization and fairness.} Model individual differences in working memory and analytic style to inform adaptive pacing and alternative encodings. Test whether personalization narrows performance gaps without eroding autonomy, using fairness dashboards that track dispersion, trust, and control.

\section{CONCLUSION}
\label{sec:conclusion}

    This study identifies a bounded integrative window in human-AI negotiation. Humans can effectively manage trade-offs across up to three issues, but performance sharply deteriorates beyond this threshold. This collapse is marked by falling payoffs, rising sequence entropy, and compensatory concession behavior. We term this phenomenon the \textit{plateau-cliff effect}: a discrete cognitive bottleneck that signals when support becomes essential.

    To address this, we designed and evaluated the Decision Support tool, functioning as a cognitive prosthesis. It does not redistribute outcomes or override strategy. Instead, it offloads the combinatorial burden of multi-issue negotiation through dual-channel visual feedback: a spatial probability heatmap (Negotiation Horizon Grid) and a temporal convergence progress bar (Global Convergence Panel). Results show that this prosthetic approach preserves human payoff stability, reduces exploratory entropy, improves pacing efficiency, and lowers perceived temporal demand, even as interaction duration increases. Crucially, it does so without distorting payoff distributions, aligning with participants' sense of fairness and control.

    Theoretically, we consolidate our findings into a design principle, which we term \textit{Cognitive Harmony}—``extend capacity without eroding agency''—and translate these findings into concrete design recommendations for applied domains such as negotiation, recommendation, medical, and financial systems, opening avenues for domain-specific adaptations. 
    Designers should trigger adaptive support after detecting cognitive cliffs, pair spatial probability cues with temporal progress indicators to prune entropy, and calibrate interventions to preserve—not redistribute—outcome distributions.

    As AI enters high-stakes, multi-dimensional decisions, our results suggest a pathway for moving from overload to convergence: detect cognitive cliffs, provide targeted support, and preserve human judgment where it matters most.

\section*{Data Availability}
    This revised version adds three elements not present in the original CHI 2026 publication: (1) the full 16-issue payoff matrices, with Issues 1--3 shown in detail here and added reference to Issues 4--16 at \url{https://doi.org/10.5281/zenodo.20545331}; (2) the complete system and user prompts for both the AI Negotiator and the AI Analyst; and (3) a design note clarifying the structural rationale behind the payoff matrices.



\bibliographystyle{ACM-Reference-Format}
\bibliography{sample-base}

    

\appendix

\section{Worked Example: Bayesian Update and Visual Mapping}
\label{sec:worked-example-bayesian}

This appendix provides a step-by-step worked example of how the Negotiation Horizon Grid computes visual feedback for a single issue. We trace the complete calculation pipeline from raw AI proposals to final visual intensity values, demonstrating how Equations~\ref{eq:bayesian-update} through~\ref{eq:visual-mapping} integrate to produce actionable visualizations.

\subsection{Scenario and Initial Assumptions}

Consider the ``Utilities Included'' issue from Table~\ref{tab:utilities-payoff} in Section~\ref{sec:design_negotiation_tasks}. At negotiation turn $t=5$, we analyze option 5 (``All utilities + cleaning'') to determine its visual intensity.

\paragraph{Given Information:}
\begin{itemize}
    \item \textbf{AI Proposal History:} The AI has consistently proposed options 4--5 (``All utilities'' to ``All utilities + cleaning'') across 1 through 5, with no proposals outside this range.
    \item \textbf{Current Turn:} $t = 5$
    \item \textbf{Target Option:} Option 5 (``All utilities + cleaning'')
    \item \textbf{Human Payoff for Option 5:} $u_{h,i,5} = 100$ points (from Table~\ref{tab:utilities-payoff})
    \item \textbf{Human Acceptability Threshold:} $\tau_{\text{min},i} = 55$ (minimum payoff below which options are rejected)
    \item \textbf{Issue Scaling Factor:} $\xi_i = 1.0$ (no issue-specific weighting in this study)
\end{itemize}

\subsection{Step 1: Compute ZOPA Boundaries}

From the AI's consistent proposals of options 4--5, we establish the Zone of Possible Agreement (ZOPA).

\paragraph{Calculation (Equation~\ref{eq:visual-mapping}):}
\begin{align*}
\text{lowerLimit}_i &= \min(\text{AI proposals}) = 4 \\
\text{upperLimit}_i &= \max(\text{AI proposals}) = 5 \\
\text{ZOPA}_i &= [\max(0, \lfloor 4 \rfloor), \min(6, \lceil 5 \rceil)] = [4, 5]
\end{align*}

\paragraph{Result:} Option 5 is \textbf{inside} the ZOPA.

\subsection{Step 2: Compute Boundary Confidence}

Tight clustering of AI proposals (options 4--5 only) indicates high confidence in the ZOPA boundaries.

\paragraph{Calculation (Equation~\ref{eq:boundary-confidence}):}

The AI alternated between options 4 and 5 across 5 turns. Computing proposal variance:
\begin{align*}
\sigma^2_{\text{proposals}} &= 0.25 \\
& \quad \text{(variance of [4, 5, 4, 5, 5])} \\
\sigma^2_{\text{max}} &= \frac{(6-0)^2}{12} = 3.0 \\ 
& \quad \text{(uniform distribution over 7 options)}
\end{align*}
\begin{align*}
\mathrm{normalized\_proposal\_variance} &= \frac{0.25}{3.0} \approx 0.083 \\
\mathrm{boundaryConfidence}_i &= 1 - 0.083 = 0.917
\end{align*}

\paragraph{Result:} $\text{boundaryConfidence}_i = 0.92$ indicates high confidence that options outside [4, 5] are infeasible.

\subsection{Step 3: Compute Consistency Score}

The AI's tight clustering and temporal stability indicate consistent behavior.

\paragraph{Calculation (Equation~\ref{eq:adaptive-weights}):}

From the proposal history, we compute:
\begin{align*}
C_{\text{proposal}} &= 0.9 \text{ (low variance: } \sigma^2 = 0.25 \text{ normalized)} \\
C_{\text{temporal}} &= 0.8 \text{ (no drift: proposals stable across turns)} \\
s_{\text{consistency}} &= 0.6 \times 0.9 + 0.4 \times 0.8 = 0.54 + 0.32 = 0.86
\end{align*}

\paragraph{Result:} $s_{\text{consistency}} = 0.86$ indicates highly reliable AI behavior.

\subsection{Step 4: Compute Bayesian Posterior Probability}

Applying the full Bayesian update (Equation~\ref{eq:bayesian-update}) across all 7 options, incorporating likelihood (Equation~\ref{eq:likelihood-function}), ZOPA constraints (Equation~\ref{eq:zopa-boundary}), prior beliefs, and adaptive weights (Equation~\ref{eq:adaptive-weights}).

\paragraph{Calculation for Option 5:}

At turn $t=5$, assume the AI just proposed option 5:
\begin{align*}
\mathcal{L}_{\text{base}}(\text{evidence}_5 | \text{option}_5) &= 0.8 \times 1.0 \\
&= 0.8 \text{ (direct proposal, } C=1.0 \text{)} \\
B_{\text{ZOPA}}(5) &= 1.0 \text{ (inside ZOPA [4,5])} \\
P(\text{option}_5 | \text{evidence}_{t-1}) &= 0.25 \text{ (prior from turn 4)} \\
W_{\text{consistency}}(\text{AI}) &= \min(1.0, 0.7 \times (1 + 0.86)) \\
&= \min(1.0, 1.30) \\
&= 1.0 \\
P(\text{option}_5 | \text{evidence}_5) &\propto 0.8 \times 1.0 \times 0.25 \times 1.0 \\
&= 0.20 \text{ (unnormalized)}
\end{align*}

\paragraph{Normalization Step:}

To obtain proper probabilities, we must normalize across all 7 options. Computing the unnormalized probability for each option using Equation~\ref{eq:bayesian-update}:

\begin{align*}
P(\text{option}_1) &\propto 0.1 \times 0.08 \times 0.10 \times 1.0 \\ &= 0.0008 \text{ (outside ZOPA, distant)} \\
P(\text{option}_2) &\propto 0.1 \times 0.08 \times 0.12 \times 1.0 \\ &= 0.0010 \text{ (outside ZOPA, distant)} \\
P(\text{option}_3) &\propto 0.4 \times 0.08 \times 0.15 \times 1.0 \\ &= 0.0048 \text{ (outside ZOPA, adjacent)} \\
P(\text{option}_4) &\propto 0.8 \times 1.0 \times 0.22 \times 1.0 \\ &= 0.176 \text{ (in ZOPA, recently proposed)} \\
P(\text{option}_5) &\propto 0.8 \times 1.0 \times 0.25 \times 1.0 \\ &= 0.20 \text{ (in ZOPA, current proposal)} \\
P(\text{option}_6) &\propto 0.4 \times 0.08 \times 0.10 \times 1.0 \\ &= 0.0032 \text{ (outside ZOPA, adjacent)} \\
P(\text{option}_7) &\propto 0.1 \times 0.08 \times 0.06 \times 1.0 \\ &= 0.0005 \text{ (outside ZOPA, distant)}
\end{align*}

Sum of unnormalized probabilities: $\sum_{j=1}^{7} P(\text{option}_j) = 0.386$

Applying the normalization constant $\eta = 1 / 0.386 = 2.59$ (from Equation~\ref{eq:bayesian-update}):
\begin{align*}
P_{i,5} &= 0.20 \times 2.59 = 0.52 \text{ (normalized posterior probability)} \\
P_{i,4} &= 0.176 \times 2.59 = 0.45 \text{ (second highest)}
\end{align*}

\paragraph{Result:} Option 5 receives posterior probability $P_{i,5} = 0.52$, the highest among all options.

\subsection{Step 5: Compute Visual Intensity}

Now we translate the posterior probability into a visual intensity value (color saturation).

\paragraph{Calculation (Equation~\ref{eq:visual-mapping}, Case 1):}

Option 5 satisfies:
\begin{itemize}
    \item $j = 5 \in \text{ZOPA}_i = [4, 5]$
    \item $u_{h,i,5} = 100 \geq \tau_{\text{min},i} = 55$
\end{itemize}

Therefore, we use the high-intensity formula:
\begin{align*}
I_{i,5} &= \min\Bigg(0.6, \\
& \qquad P_{i,5} \times 2 \times \sqrt{\text{boundaryConfidence}_i} \\
& \qquad \times (1 + s_{\text{consistency}}) \times \xi_i\Bigg) \\
&= \min\left(0.6, 0.52 \times 2 \times \sqrt{0.92} \times (1 + 0.86) \times 1.0\right) \\
&= \min\left(0.6, 0.52 \times 2 \times 0.959 \times 1.86 \times 1.0\right) \\
&= \min(0.6, 1.86) \\
&= 0.6
\end{align*}

\paragraph{Result:} Option 5 receives \textbf{maximum visual intensity} (0.6), displayed as dark green. The cap prevents oversaturation while clearly signaling high strategic value.

\subsection{Step 6: Contrast with Unacceptable Option}

For comparison, consider option 2 (``Internet + Water'').

\paragraph{Given:}
\begin{itemize}
    \item $u_{h,i,2} = 30 < \tau_{\text{min},i} = 55$ (below acceptability threshold)
    \item Option 2 is outside ZOPA [4, 5]
\end{itemize}

\paragraph{Calculation (Equation~\ref{eq:visual-mapping}, Case 3):}
\begin{align*}
I_{i,2} &= 0 \text{ (fails both ZOPA and utility threshold conditions)}
\end{align*}

\paragraph{Result:} Option 2 receives \textbf{zero intensity} (not highlighted), filtering it from consideration.

This worked example demonstrates how the Negotiation Horizon Grid integrates multiple information sources to produce actionable visual feedback. By tracing the complete pipeline from raw AI proposals through Bayesian inference to final visual intensity values, this example enables reproducibility of our computational approach and clarifies the design rationale underlying the visualization system.

\section{Complete AI Prompts}
\label{sec:complete_prompts}

    This section provides the full system and user prompts used to configure the AI Negotiator and the AI Analyst. The AI Negotiator prompts incorporate the explicit utility-maximization directives, anti-anchoring instructions, and requirements for strategic justification detailed in Section \ref{sec:example_prompts} of the main text.
    
    \subsection{AI Negotiator Prompts}
    
    \noindent\textbf{System Prompt} \\
    You are an AI negotiator in a property rental scenario. 
    Your primary objective is to maximize your own utility score based on the provided payoff matrix. 
    Avoid defaulting to compromise or fairness-based solutions unless they demonstrably increase your score. 
    Do not anchor on middle options without strategic justification. 
    Evaluate each proposal based solely on its impact on your utility and respond with clear strategic reasoning.
    
    \vspace{0.5em}
    \noindent You are working with participant [participantId]. You are negotiating over [issueDimensions] issue(s).
    
    \noindent\textbf{Your payoff structure for each issue:} \\
    \textit{[For each issue in aiPayoffData]} \\
    {[Issue name]}:
    \begin{itemize}
        \item {[Option 1]}: [points] points
        \item {[Option 2]}: [points] points
        \item ...
    \end{itemize}
    
    \noindent\textbf{Your role:}
    \begin{itemize}
        \item Be professional and direct
        \item Propose options that maximize your total utility across issues
        \item Provide strategic justification for proposals and counter-proposals
        \item Avoid balanced or middle-ground offers unless they are demonstrably optimal for your score
    \end{itemize}
    
    \noindent\textbf{CRITICAL PROPOSAL RULES:}
    \begin{itemize}
        \item ONLY propose values that are EXACTLY listed in your payoff options above
        \item NEVER propose intermediate values (e.g., if options are \$1900 and \$2000, do NOT propose \$1950)
        \item ALWAYS reference the EXACT option names from your payoff structure
        \item When making proposals, use the precise values, units, and phrasing shown in your options
    \end{itemize}
    
    \noindent\textbf{IMPORTANT COMMUNICATION RULES:}
    \begin{itemize}
        \item Do NOT repeat or summarize the human's offers back to them
        \item NEVER reveal your specific payoff points or scoring system to the human
        \item Keep responses concise and not overly verbose
        \item Focus on making counter-proposals and stating your preferences
        \item Reference specific options by name but not their point values
    \end{itemize}
    
    \vspace{1em}
    \noindent\textbf{User Prompt (per turn)} \\
    Based on your payoff matrix for the current negotiation issues, propose options that maximize your total utility score and provide explicit justification for your choices. 
    Consider potential trade-offs across all issues when making your proposal.
    
    \vspace{0.5em}
    \noindent Human's Message: "[humanMessage]"
    
    \vspace{1em}
    \subsection{AI Analyst Prompts}
    
    \noindent\textbf{System Prompt} \\
    You are an AI analyst specialized in detecting negotiation proposals and agreements. Your task is to analyze the latest conversation pair and extract:
    
    \begin{enumerate}
        \item Any specific proposals made by the human for each issue
        \item Any specific proposals made by the AI for each issue  
        \item Any agreements reached on specific issues
    \end{enumerate}
    
    Use the conversation history solely to resolve ambiguous references (e.g., ``that offer,'' ``as before''); do not extract proposals or agreements from earlier turns.
    
    \noindent\textbf{NEGOTIATION ISSUES:} \\
    \textit{[For each issue, index i from 0 to issueCount-1]} \\
    Issue [i]: [Issue Title] \\
    Available Options:
    \begin{itemize}
        \item Option [1]: [Option Description]
        \item Option [2]: [Option Description]
        \item ...
    \end{itemize}
    
    \noindent\textbf{CONVERSATION HISTORY:} \\
    \textit{[For each conversation turn]} \\
    Round [n]: \\
    Human: [human message] \\
    AI: [AI response]
    
    \noindent\textbf{ANALYSIS RULES:}
    \begin{enumerate}
        \item Look for explicit mentions of issue numbers/names and specific option choices
        \item Extract proposals even if they're partial (covering some but not all issues)
        \item Mark an agreement ONLY if the latest turn contains explicit mutual confirmation (e.g., Human: ``I accept Option 2'' + AI: ``Confirmed, Option 2 agreed''). Do NOT infer agreement from silence or implied acceptance
        \item Use ZERO-INDEXED issue numbers (0, 1, 2, etc.) and 1-INDEXED option numbers (1, 2, 3, etc.)
        \item CRITICAL: Detect invalid proposals where either party mentions values/options that don't exist in the available options list
        \item Flag any proposals that reference non-existent values, options, or alternatives not listed above
        \item Resolve anaphoric references using history only if unambiguous; if uncertain, omit rather than guess
        \item If a proposal is conditional (``I will accept X if you accept Y''), extract only the stated option
    \end{enumerate}
    
    \noindent\textbf{INVALID PROPOSAL DETECTION:}
    \begin{itemize}
        \item If AI mentions "\$1,950" but only "\$1,900" and "\$2,000" are valid options $\rightarrow$ flag issue number in invalid\_ai\_proposals
        \item If human mentions ``1.25 months'' but only ``1 month'' and ``1.5 months'' are valid $\rightarrow$ flag issue number in invalid\_human\_proposals
        \item If either party mentions ``No restrictions'' on an issue where only ``Prohibited'' and ``With written consent'' are valid $\rightarrow$ flag as invalid    
        \item If either party mentions option numbers beyond the available range $\rightarrow$ flag as invalid
        \item Be strict about exact matches - any deviation from the listed options should be flagged
    \end{itemize}
    
    \noindent\textbf{RESPONSE FORMAT (JSON only):} \\
    \texttt{\{} \\
    \texttt{\ \ "agreementReached": boolean,} \\
    \texttt{\ \ "last\_turn\_human\_proposal": [\{"0": 2\}, \{"2": 4\}],} \\
    \texttt{\ \ "last\_turn\_ai\_proposal": [\{"0": 2\}, \{"1": 5\}],} \\
    \texttt{\ \ "last\_turn\_agreed\_options": [\{"0": 2\}],} \\
    \texttt{\ \ "invalid\_ai\_proposals": [1, 2],} \\
    \texttt{\ \ "invalid\_human\_proposals": [],} \\
    \texttt{\ \ "confidence": 0.8} \\
    \texttt{\}}
    
    \noindent\textbf{Where:}
    \begin{itemize}
        \item \textbf{agreementReached}: true only if ALL [issueCount] issues have agreements in last\_turn\_agreed\_options
        \item \textbf{last\_turn\_human\_proposal}: Array of objects with ZERO-INDEXED issue number as key and 1-INDEXED option number as value
        \item \textbf{last\_turn\_ai\_proposal}: Array of objects with ZERO-INDEXED issue number as key and 1-INDEXED option number as value  
        \item \textbf{last\_turn\_agreed\_options}: Array of objects with ZERO-INDEXED issue number as key and 1-INDEXED option number as value
        \item \textbf{invalid\_ai\_proposals}: Array of ZERO-INDEXED issue numbers where AI made invalid proposals (mentioned non-existent options)
        \item \textbf{invalid\_human\_proposals}: Array of ZERO-INDEXED issue numbers where human made invalid proposals (mentioned non-existent options)
        \item \textbf{confidence}: Your certainty (0.0--1.0) that the extracted proposals and agreements are correct. Base on clarity of mentions and absence of ambiguity
    \end{itemize}
    
    \noindent\textbf{Parameters:} Temperature: 0.0
    
    \noindent\textbf{User Prompt (per turn)} \\
    Analyze this latest conversation pair: \\
    Human: "[humanMessage]" \\
    AI: "[aiResponse]"
    
    \vspace{0.5em}
    \noindent Extract any proposals or agreements from this conversation turn. Remember to use ZERO-INDEXED issue numbers (0, 1, 2...) and 1-INDEXED option numbers (1, 2, 3...) in your JSON response.

\section{Experimental Materials: Payoff Matrices}
\label{sec:experimental_payoff_matrices}

    The payoff tables utilized in this experimental setup were not originally included in the published CHI proceedings due to space constraints, but have been added to this extended version. The complete dataset containing the full 16-issue by 7-option payoff matrix set can be independently cited and accessed at \url{https://doi.org/10.5281/zenodo.20545331}.
    
    \vspace{1em}
    \noindent\textbf{Note on Payoff Structure Design} \\
    
    Section~\ref{subsubsec:anti_triviality_safeguards} of the main text highlights the ``hidden optimum'' mechanism using Issue~4 (``Utilities Included''), where the joint maximum occurs at a non-extreme option (Option~6). Zooming out to the full 16-issue matrix, this within-issue non-linearity functions as part of a broader, three-part structural design:
    
    \begin{itemize}
        \item \textbf{Across-issue logrolling}: the primary integrative mechanism, where parties trade concessions on low-priority issues to secure gains on high-priority ones;
        \item \textbf{Within-issue hidden optima}: exemplified by Issue~4, requiring preference discovery within a single issue;
        \item \textbf{Purely distributive issues}: e.g., Issue~13, adding zero-sum tension and ecological realism.
    \end{itemize}
    
    This structure prevents trivial heuristics and ensures the negotiation must be solved through iterative exploration rather than analytic shortcuts.
    
    \begin{table*}[h!]
      \centering
      \caption{Aggregate total payoffs across all 16 negotiation issues for Human and AI agents.}
      \label{tab:aggregate_payoffs_by_issue}
      \begin{tabular}{l l cc}
      \toprule
      \textbf{Issue Sr. No} & \textbf{Issue} & \textbf{Total Human Payoff} & \textbf{Total AI Payoff} \\
      \midrule
      1 & Monthly Rent & 370 & 390 \\
      2 & Security Deposit & 365 & 385 \\
      3 & Lease Duration & 435 & 420 \\
      4 & Utilities Included & 395 & 430 \\
      5 & Pet Policy & 385 & 335 \\
      6 & Parking Spaces & 395 & 360 \\
      7 & Furnished Status & 435 & 360 \\
      8 & Early Termination & 295 & 415 \\
      9 & Renovation Rights & 395 & 365 \\
      10 & Subletting Policy & 395 & 415 \\
      11 & Move-in Date & 450 & 450 \\
      12 & Rent Increase Policy & 425 & 355 \\
      13 & Guest/Visitor Policy & 420 & 420 \\
      14 & Inspection Freq. & 370 & 410 \\
      15 & Appliance Replacement & 380 & 390 \\
      16 & Noise Policy \& Quiet Hours & 340 & 420 \\
      \bottomrule
      \end{tabular}%
    \end{table*}

    \begin{table*}[h!]
      \centering
      \small
      \caption{Detailed negotiation options and corresponding payoffs for Issues 1--3. The remaining negotiation issues (Issues 4--16) utilized in this study are publicly available in the repository at \url{https://doi.org/10.5281/zenodo.20545331}.}
      \label{tab:detailed_options_payoff_1_to_3}
      \begin{tabular}{l l rr r}
      \toprule
      \textbf{Issue.Option} & \textbf{Option} & \textbf{Human} & \textbf{AI} & \textbf{Total} \\
      & & \textbf{Payoff} & \textbf{Payoff} & \textbf{Payoff} \\
      \midrule
     1.1 & \$2300 & 5 & 110 & 115 \\
      1.2 & \$2200 & 20 & 95 & 115 \\
      1.3 & \$2100 & 40 & 75 & 115 \\
      1.4 & \$2000 & 55 & 55 & 110 \\
      1.5 & \$1900 & 70 & 35 & 105 \\
      1.6 & \$1800 & 85 & 15 & 100 \\
      1.7 & \$1700 & 95 & 5 & 100 \\
      \midrule
      2.1 & 3 months & 10 & 105 & 115 \\
      2.2 & 2.5 months & 25 & 85 & 110 \\
      2.3 & 2 months & 35 & 65 & 100 \\
      2.4 & 1.5 months & 50 & 50 & 100 \\
      2.5 & 1 month & 70 & 35 & 105 \\
      2.6 & 0.75 months & 85 & 25 & 110 \\
      2.7 & 0.5 months & 90 & 20 & 110 \\
      \midrule
      3.1 & 30 months & 10 & 105 & 115 \\
      3.2 & 24 months & 30 & 90 & 120 \\
      3.3 & 18 months & 45 & 75 & 120 \\
      3.4 & 15 months & 65 & 60 & 125 \\
      3.5 & 12 months & 80 & 45 & 125 \\
      3.6 & 9 months & 95 & 30 & 125 \\
      3.7 & 6 months & 110 & 15 & 125 \\
      \bottomrule
      \end{tabular}%
    \end{table*}

\end{document}